\documentclass[journal]{IEEEtran}

\usepackage{amsthm}
\usepackage{amssymb}
\usepackage{amsfonts}
\usepackage{amsfonts,subfigure,multicol,color,verbatim,graphicx,cite,epsfig,amssymb,amsmath,cases,bm,xcolor,multirow,array}
\usepackage{algorithm,algorithmicx,algpseudocode}
\usepackage{multirow}
\usepackage{multicol}
\usepackage{supertabular}
\usepackage{array}
\usepackage{colortbl}
\usepackage{longtable}
\usepackage{subfigure}
\usepackage{dsfont}
\usepackage{url}

\usepackage{epstopdf,booktabs}

\allowdisplaybreaks[4]

\usepackage{soul}
\usepackage{color, xcolor}
\soulregister{\cite}7
\soulregister{\citep}7
\soulregister{\citet}7
\soulregister{\ref}7
\soulregister{\pageref}7
\soulregister{\eqref}7

\UseRawInputEncoding
\begin{document}

\title{Timing Advance Estimation in Low Earth Orbit Satellite Networks}\vspace{1.5cm}

\author{Jianfeng~Zhu,~Yaohua~Sun~and~Mugen~Peng,~\IEEEmembership{Fellow,~IEEE}
\thanks{
Manuscript received 15 May 2023; revised 4 August 2023; accepted 26 September 2023. Date of publication 17 October 2023; date of current version 14 March 2024. This work was supported in part by the Beijing Municipal Science and Technology Project under Grant Z211100004421017, in part by the National Natural Science Foundation of China under Grants 62371071 and 62001053, and in part by the Young Elite Scientists Sponsorship Program by CAST under Grant 2021QNRC001. The review of this article was coordinated by Dr. Tomaso De Cola. (\emph{Corresponding author: Yaohua Sun.})

The authors are with the State Key Laboratory of Networking and Switching Technology, Beijing University of Posts and Telecommunica-tions, Beijing 100876, China (e-mail: jianfeng@bupt.edu.cn; sunyaohua@bupt. edu.cn; pmg@bupt.edu.cn).

Digital Object Identi?er 10.1109/TVT.2023.3325328.
}}

\markboth{IEEE TRANSACTIONS ON VEHICULAR TECHNOLOGY, VOL. 73, NO. 3, MARCH 2024}%
{Shell \MakeLowercase{\textit{et al.}}}
\maketitle

\begin{abstract}
Low earth orbit (LEO) satellite communication based on 3GPP standard is seen as a promising solution to rolling out communication services in areas without terrestrial base stations.
However, due to the fast movement of satellites and large beam footprint size, the existing 5G timing advance (TA) estimation mechanism cannot be directly applied when global navigation satellite system is unavailable.
In this article, an enhanced TA estimation approach is proposed for LEO satellite communication networks.
Specifically, a user-side time-frequency pre-compensation method is introduced at first, which leverages frequency offset measurement on synchronization signal blocks broadcasted by satellites in initial cell search phase.
For the random access phase, the upper bound of inter-preamble interference incurred by partial-period cross-correlation operations is derived for a preamble format advised by 3GPP, and it is shown that the interference level is closely related to the square of the number of such operations.
Inspired by this result, a cyclic prefix free preamble format is further designed, which features extended guard time, differential power allocation and flexible preamble structure.
Numerical results show that our proposal can reduce the missed detection rate of preamble within a beam. Particularly, the missed detection rates of preamble under 32, 48, and 64 users are lower than 1$\%$ when SNR = -6 dB, which is a significant improvement compared to baselines.
In addition, our proposal can limit the TA estimation error of the detected users to the time length of 25 time-domain sampling points when the subcarrier spacing is 30 kHz and operation frequency is 27 GHz.
\end{abstract}


%---------------------------------
\section{Introduction}
%---------------------------------
In order to provide communication services in disaster areas, deserts, oceans and polar regions, the integration of satellites into 5G and beyond has attracted significant attention~\cite{integration1,integration2,integration3,integration4,integration5,integration6}.
Currently, several specification works are going on in 3GPP~\cite{38.811}~\cite{38.821}, and one of discussions focuses on 5G random access procedure enhancement.
As a contention-based protocol, random access procedure is mainly utilized to achieve uplink synchronization between a user equipment (UE) and a base station~\cite{RA1}, where preamble format design is the key to timing advance (TA) estimation.

According to~\cite{cell_size}, when the subcarrier spacing (SCS) of physical random access channel (PRACH) is 1.25 kHz, the coverage radius of base stations in terrestrial NR can be extended to up to 100 km.
However, when satellites operate at KA band, commonly used preamble designs and detection procedures for terrestrial NR lead to considerable timing errors and access failures due to large frequency offsets~\cite{CFO_damage}.
To guarantee the accuracy of TA estimation when large frequency exists,
the maximal supported cell radius in terrestrial NR is 9.7 km, corresponding to major semi-axis of the beam cell less than 35 km when SCS of PRACH is 30 kHz~\cite{beam_size}.
Overall, the 5G preamble format proposed in 3GPP standard TS 38.221 of Rel. 17 cannot be applied directly in 5G LEO networks~\cite{38.221}, which feature large equivalent cell radius and frequency offset.

To cope with large cell radius, researchers have proposed several feasible solutions for random access, including limiting beam footprint size, reducing the equivalent cell radius based on UE positioning and designing enhanced preamble formats.
Guidotti $et$ $al.$~\cite{beam_size} present an adaptive beam footprint size design to limit the range of equivalent cell radius.
However, flexible beam size designs have high requirements on satellite antennas, and small beam footprints will also incur large overheads due to frequent inter-beam handover.
Based on Open Air Interface (OAI), authors in~\cite{GNSS_practice} develop a practical testbed to validate the feasibility of 5G preamble formats proposed in the 3GPP standard TS 38.221 of Rel. 17~\cite{38.221}, where UEs with global navigation satellite system (GNSS) capability can estimate an exact TA based on its position before starting random access procedure, which helps reduce equivalent cell radius.
In addition, satellite access technology for GNSS independent operation scenarios
is studied in 3GPP Rel. 19~\cite{R19}, where UEs lack positioning information of them.
To cope with this challenge, UE position in~\cite{location} is acquired by solving a quadratic optimization problem relying on a series of timing and frequency offset measurements based on synchronization signal blocks (SSB).
However, to attach a high level of positioning accuracy,  UEs require a large timing window to obtain measurements.
In~\cite{beam_subarea}, a two-step TA estimation method is presented, which first divides a beam into different sub-regions according to propagation delay and then two preamble formats are designed to calculate the integer and fractional parts of TA, respectively.
Authors in~\cite{zhenli} investigate a preamble format design based on one root Zadoff-Chu (ZC) sequence with different cyclic shifts, and a large cell radius is supported by jointly using all autocorrelation peaks at satellites generated by multiple ZC sequences.

Although the prior works achieve good performance, they ignore the impact of frequency offset on TA estimation accuracy.
Recently, various preamble formats that are robust to frequency offset have been proposed.
In~\cite{COE}, the frequency offset in uplink is pre-compensated at UE side using the frequency offset in downlink, and a preamble format based on repeating transmission and frequency hopping is adopted for NB-IoT scenarios.
Due to hopping transmission, it incurs significant time-frequency resource overhead for random access.
Authors in~\cite{weighted} review the framework of preamble format design based on ZC sequence weighted by pseudo-noise (PN), which can also alleviate the effect of frequency offset.
A preamble format is developed in~\cite{2NR-preamble} by concatenating two NR PRACH preambles with different roots, where each preamble consists of cyclic prefix (CP), ZC sequences, and guard time.
Similarly, the authors in~\cite{2root} propose a root pair selection method for preamble design and then generate a preamble with two different ZC sequences, where the influence of frequency offset on ZC sequences with different roots is investigated.
In~\cite{ZC_decomposed}, a ZC sequence is decomposed into multiple short subsequences to reduce the negative impact of frequency offset, where short subsequences can be regarded as new ZC sequences.
Although the preamble detection schemes of \cite{COE,weighted,2NR-preamble,2root,ZC_decomposed} are robust to frequency offset, they require a two-step TA estimation.
In~\cite{one-step1}, a preamble format with different roots is proposed, and the authors design a one-step preamble detection for TA estimation, which can exactly estimate TA in large frequency offset scenarios.
A preamble format with conjugate-symmetric ZC sequences is developed in~\cite{one-step2}, where the authors use the same preamble detection procedure with~\cite{one-step1}.
However, since the detected peak in~\cite{one-step1} and \cite{one-step2} only occurs when timing index output by preamble detection procedure aligns with the TA of received preamble, the base station is required to search for each possible timing index and hence computational complexity is high.
The above works provide multiple optional preamble formats and preamble detection procedures to support accurate TA estimation in large equivalent cell radius and frequency offset scenarios.
Nevertheless, they ideally assume that inter-preamble interference has a negligible effect on preamble detection procedures, which is unreasonable
when a large number of UEs perform random access simultaneously.

In this article, an enhanced TA estimation approach is proposed to achieve uplink synchronization in 5G LEO satellite networks.
Our proposal focuses on UEs without GNSS capability in some special contexts  proposed by 3GPP~\cite{R19}.
For example, GNSS signals can be heavily interfered in military scenarios and the navigation module of a fixed UE may be damaged unexpectedly.
The meaning of UE without GNSS capability represents that UEs cannot obtain a high level of accuracy of positioning information from current navigation and positioning systems, including the Global Positioning System and the BeiDou Navigation Satellite System.
However, UEs can achieve self-positioning based on SSBs broadcasted from satellites.
The proposal is illustrated for satellites with regenerative payloads, but it should be highlighted that the proposal can also be applied to transparent satellites if the gateway station on the ground performs time-frequency compensation for feeder links.
Motivated by recently advances in multiple satellites cooperative communication research \cite{Multisatellite}, UEs are assumed to simultaneously receive SSBs from multiple satellites and obtain ephemerides.
With such capability, a positioning-based coarse time-frequency pre-compensation method is adopted before random access procedure in our proposal, where each UE only detects SSBs from three satellites for only one time.
The central idea of our proposal is to limit the residual frequency offset of the uplink to less than half of the SCS  of PRACH.
Meanwhile, residual timing offsets after time pre-compensation are still relatively large, which requires an enhanced preamble format and detection procedure.
Then, we propose a preamble format design compatible with 3GPP
standard based on inter-preamble interference analysis, whose detection procedure can fully reuse current preamble detection function modules in 3GPP
standard.
With the proposed preamble format, satellites can achieve high accuracy of TA estimation in multi-UE case.
The main contributions of this article are summarized as follows.
\begin{itemize}
\item
To tackle the challenges in TA estimation incurred by large equivalent cell radius and frequency offset, we propose a time-frequency pre-compensation method for UEs without GNSS capability before preamble transmission.
In this method, initial TA, doppler shift and frequency offset caused by local oscillators are estimated at UE side by detecting SSBs from multiple satellites, contributing to reducing the equivalent cell radius and uplink frequency offset.

\item
Given that inter-preamble interference can be classified into full-period cross-correlation interference and partial-period cross-correlation interference and the latter dominates the total interference, its theoretical upper bound is derived to guide preamble design, which indicates that it is beneficial to reduce the negative impact of partial-period cross-correlation operations and increase the power difference between interfering ZC sequence and the desired ZC sequence at the satellite.

\item
Motivated by the above analysis, a flexible preamble format is proposed for UEs without GNSS capability in LEO satellite networks, featuring the flexible cascading in time domain and differential power allocation of ZC sequences with different roots, and it is proved that flexible cascading strategy can lower the negative impact of partial-period cross-correlation.
By simulation, the missed detection rate of our preamble design significantly decreases compared to other baselines, which contributes to high TA estimation accuracy and more supported UEs in random access.
Particularly, the TA estimation error of the detected UEs does not exceed $10\%$ of normal CP length under 30 kHz subcarrier spacing at operation frequency of 27 GHz.
Moreover, the proposed preamble format design can also be applied to low frequency band, such as L, S and X band.
\end{itemize}

The remainder of this paper is organized as follows.
Section~\ref{sec:system_model} provides system scenario, the general procedure of preamble detection and TA estimation, and an overview of our proposed enhanced TA estimation approach.
In Section~\ref{sec:time-frequency pre-compensation}, a time-frequency pre-compensation method is developed, and inter-preamble interference analysis, preamble detection error analysis, and proposed preamble format design are illustrated in Section~\ref{sec:preamble_design1}. Finally, simulation and analysis are presented in Section~\ref{sec:sim} followed by the conclusion
in Section~\ref{sec:con}.
The notation used in this paper is summarized in Table~\ref{tab:notation}.

\begin{table}[!t]
\centering \caption{Notation}
\label{tab:notation}
\scriptsize
\begin{tabular}{l l} \toprule
\textbf{Notation} & \textbf{Definition} \\
\midrule
$\mathcal{U}$ & The set of UEs \\
$\mathcal{S}$ & The set of satellites \\
$T^u$ & The actual TA of UE $u$\\
$T_{zc}$ & The duration of a ZC sequence in time domain \\
$K_i^u$ & The integer part of TA estimation \\
$K_f^u$ & The fractional part of TA estimation \\
$T_d$ & The maximal differential TA within a beam\\
$N_{zc}$ & The length of a ZC sequence \\
$N_{idft}$ & The sampling point number of an OFDM  symbol  in  time  \\& domain \\
$\{x(n,r_i^u)\}$ & A ZC sequence with root $r_i^u$ \\
${PDP(m,r_i^u)}$ &The sequence is generated by PDP computation   between \\& a sequence and  sequence $\{x(m,r_i^u)\}$\\
$T_s$ & The time domain sampling point interval \\
$\tau_u$ & The arrival time of UE $u$'s preamble normalized  by $T_s$\\
$Z_l$ & The number of ZC sequence in a preamble \\
$\{y_{u,j}^\prime(n)\}$ & The detected $j$-th subsequence of UE $u$ 's preamble in \\&module 2 of the  general TA  estimation procedure\\
$b^u$ & The position of the ZC sequence with root $r_b^u$ in  the \\& preamble  of UE $u$\\
$T_{pre}^u$ & The time pre-compensation value of UE $u$ \\
$(x_s,y_s,z_s)$ & The position of satellite $s$ in ECEF \\
$(x_u,y_u,z_u)$ & The position of UE $u$ in ECEF \\
$(R, \theta_u, \varphi_u )$ & The position of a geostationary UE $u$ in spherical\\& coordinate system \\
$(v_s^x,v_s^y,v_s^z)$ & The velocity of satellite $s$ in ECEF  \\
$f_s$ & The operation frequency of satellite $s$ \\
$f_s^u$ & The doppler shift between UE $u$ and satellite $s$ \\
$f_{lo}^u$ &  The frequency offset between UE $u$ and satellite $s$  \\& caused by local oscillators\\
$f_{u,s}^{down}$ & The downlink CFO between UE $u$ and satellite $s$\\
$f_{u,s}^{up}$ & The uplink CFO between UE $u$ and satellite $s$ \\
$\alpha^{u^\prime}$ & The start point of the interfering sequence from UE $u^\prime$ \\
$\beta^{u^\prime}$ & The length of the interfering sequence from UE $u^\prime$ \\
$\{\rho_{\alpha^{u^\prime},\beta^{u^\prime}}(n)\}$  & A windowing sequence to represent the overlapping  \\& area of the interfering sequence and $\{x(n,r_i^{u^\prime})\}$ \\
$\widehat \rho_{\alpha^{u^\prime},\beta^{u^\prime}}(n)$ & The frequency response of $\rho_{\alpha^{u^\prime},\beta^{u^\prime}}(n)$ \\
$M_2(\beta^{u^\prime},k)$ & The upper bound of inter-preamble interference  in Case 2 \\
$M_3(\beta^{u^\prime},k_1,k_2)$ & The upper bound of inter-preamble interference in Case 3 \\
$R(k,\beta,k_1,k_2)$ & The ratio of  $M_3(\beta,k_1,k_2)$ to autocorrelation peak \\
$Pr_u(j)$ & The probability of the j-th subsequence ${y_{u,j}(n)}$ \\&performs  partial-period cross-correlation operation  \\& under the given cascading order of ZC sequences\\
$Pr_u^\prime(j)$ & The probability of the j-th subsequence ${y_{u,j}(n)}$ \\& performs  partial-period cross-correlation operation \\& under the flexible cascading order of ZC sequences \\
\bottomrule
\end{tabular}
\end{table}

%---------------------------------
\section{SYSTEM MODEL}\label{sec:system_model}
%---------------------------------

%---------------------------------
\subsection{Scenario description}\label{sec:preamble_option}
%---------------------------------

\begin{figure}[hbpt]
	\center
	\includegraphics[width=2.5in]{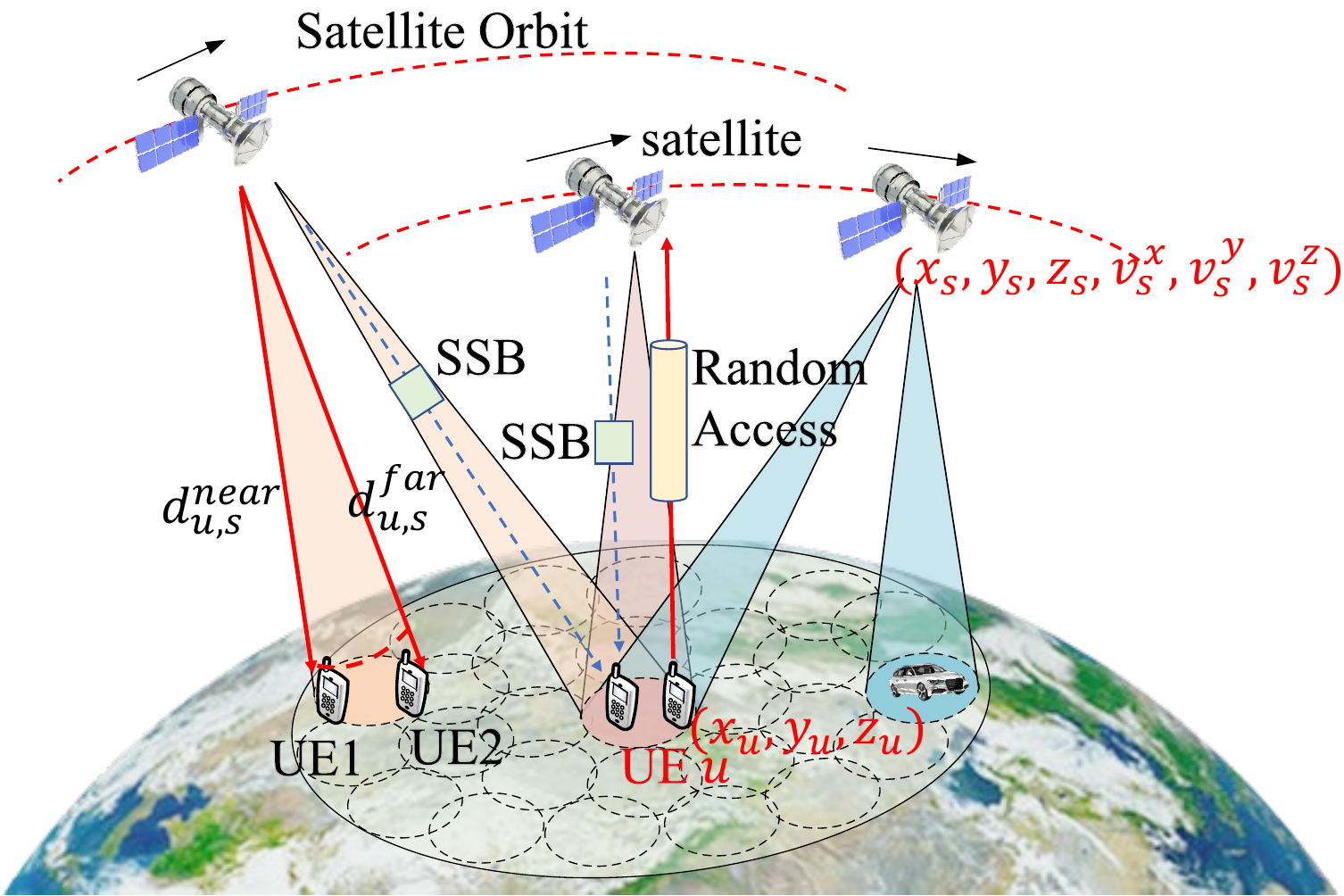}
	\caption{An LEO satellite network scenario.}\label{fig1}
\end{figure}

As indicated in Fig.~\ref{fig1}, each UE is assumed to receive SSBs broadcasted by multiple satellites with base station functions onboard in the initial cell search phase~\cite{Multisatellite}.
Since SSB can help UEs detect system information of a satellite cell and ephemeris data, hence, UE can derive position coordinates and velocity of satellites before attempting to access the cell~\cite{38.821}.
System information further includes operation frequency, available random access preambles, as well as optional random access channel occasions (RO).
With RO information, UEs transmit preambles and uplink payload at the reserved time-frequency resource blocks.
Then, satellites detect preambles and estimate TA for each UE to achieve uplink synchronization.
Assume that satellites operate in time division duplex (TDD) mode, and the benefit lies in that satellites can flexibly adjust the ratio of uplink and downlink slots according to service demands to improve spectrum efficiency.
Since inter-beam interference can be mitigated by proper frequency reuse and hopping strategy among beams, we focus on UEs accessing a typical beam in the analysis below.

Denote the set of UEs as $\mathcal{U}=\{1,2,...,U\}$.
Define $d_{u,s}$ as the distance between UE $u$ and satellite $s$, and propagation delay between satellite $s$ and UE $u$ is $T_{u,s} = d_{u,s}/c$,  where $c$ is the speed of light.
Denote $d_{u,s}^{far}$,  $d_{u,s}^{near}$ as the longest and shortest distance from a point in the beam cell to satellite $s$, respectively.
The difference between them is the equivalent cell radius.
The maximal and minimal propagation delay of UEs in a beam cell are calculated as $T_{u,s}^{max} = d_{u,s}^{far}/c, T_{u,s}^{min} = d_{u,s}^{near}/c$, respectively.
Consider that TA is twice the propagation delay, then the actual TA of UE $u$ can be expressed as
\begin{equation}
	T^u=2T_{u,s}=2T_{u,s}^{min}+K_i^u T_{zc} + K_f^u,
	\label{eq:TA repre}
\end{equation}
where $T_{zc}$ is the duration of a ZC sequence in time domain, $K_i^u$ is a non-negative integer and $0\leq K_f^u < T_{zc}$.
In the following, $K_f^u$ and $K_i^u $ are called the fractional part and the integer part of TA estimation, respectively.
Note that the coverage of a satellite beam is much larger than that of terrestrial cells, which leads to a considerable difference in $T^u$ among UEs within the beam.
Denote the maximal differential TA among UEs as $T_d$, which is given by
\begin{equation}
	\begin{aligned}
		T_d= 2(T_{u,s}^{max}-T_{u,s}^{min}).
	\end{aligned}
	\label{eq:differential_TA}
\end{equation}
Particularly, if the beam radius is 100 km, $T_d$ can achieve several milliseconds.
Considering that UEs without GNSS capability have little knowledge about $T^u$ in initial random access phase, the arrival time of UE's preamble at the satellite is quite uncertain, which requires preamble format to support estimating TA in a large duration.
In addition, the preamble missed detection rate and TA estimation error increases rapidly when the frequency offset is larger than half of SCS of PRACH~\cite{hua1}.
Specifically, the carrier frequency offset (CFO) between UEs and satellites consists of doppler shift and frequency offset caused by local oscillators~\cite{COE}.
According to \cite{chen_shanzhi}, doppler shift can reach 720 kHz at 30 GHz when satellite altitude is 600 km, and maximum SCS that 5G can support is 240 kHz.
In addition, the authors of~\cite{damange_O} point out that the frequency offset of local oscillators may be larger than the SCS of PRACH.
Meanwhile, the existing frequency offset estimation algorithm can accurately estimate the frequency offset of the downlink, but it cannot distinguish the frequency offset caused by the doppler effect from the crystal oscillator shift~\cite{damange_LO}.
Overall,  we cannot ignore the frequency offset caused by local oscillators.

Traditional preamble formats consist of CP, preamble and guard time~\cite{38.221}.
The preamble is generated by ZC sequences cascaded in time domain, while the involvement of CP and guard time is to reduce inter-symbol interference.
In 3GPP\cite{38.821}, several optional enhanced preamble formats based on ZC sequences are considered for UEs without GNSS capability, which are given below.

1) Option-1: preamble constructed by a single ZC sequence with larger SCS and repetition number;

2) Option-2: preamble constructed by scrambled ZC sequences;

3) Option-3: preamble constructed by ZC sequences with different roots.

In the following, we provide basic symbolic representations of preamble made up by ZC sequences with two roots, which is based on Option-3.
The most popular form of ZC sequences is written as
\begin{equation}
	x(n,r) = e^{- j \frac{\pi rn(n+1)}{N_{zc}}},
	\label{eq:zc_expression}
\end{equation}
where $N_{zc}$ is the sequence length, $n=0,1,...,N_{zc}-1$ represents the element index in sequence, $r$ is the root of ZC sequence.
Define $r_a^u$ and $r_b^u$ as the roots of ZC sequences adopted by UE $u$, as shown in Fig.~\ref{fig2}(a).
Denote the number of ZC sequences in a preamble as $Z_l$ and an integer variable $i \in [1,Z_l]$.
Denote $r_i^u$ as the root of the $i$-th ZC sequence in the preamble, and $r_i^u\in \{r_a^u, r_b^u \}$.
Define $N_{idft}$ as the sampling point number of an OFDM symbol in time domain, which depends on the size of SCS.
After performing $N_{zc}$ point discrete Fourier transform (DFT), frequency mapping and $N_{idft}$ point inverse discrete Fourier transform (IDFT), the $i$-th time-domain ZC sequence $\{x_i^u(l)\}$ transmitted by UE $u$ with $l=1,2,...,N_{idft}$ is given by\cite{38.221}
\begin{equation}
	x_i^u(l)=P_i^u\sum_{g=1}^{N_{idft}} e^{j\frac{2\pi gl}{N_{idft}}}\sum_{n=0}^{N_{zc}-1}x(n,r_i^u)e^{-j\frac{2\pi
			gn}{N_{zc}}},
	\label{eq:transmitted signal}
\end{equation}
where $P_i^u$ is the time-domain signal amplitude.

Denote the normalized uplink frequency offset relative to the SCS of PRACH as $f_u$.
Then, the $i$-th time-domain ZC sequence $\{z_i^u(l)\}$ from UE $u$ received at the satellite can be written as
\begin{equation}
	z_i^u(l)=A_uP_i^u e^{j2\pi f_u T_s l}x_i^u(l-\tau_u) +w_i(l),
	\label{eq:gNBreceived signal}
\end{equation}
where $A_u$ denotes uplink channel attenuation, $T_s$ is the sampling point interval when satellite $s$ samples PRACH in time domain, $w_i(l)$ is the additive white Gaussian noise with zero mean and variance $\sigma_i^2$, and $\tau_u$ is the normalized arrival time of UE $u$'s preamble, which is given by
\begin{equation}
	\tau_u = \lceil \frac{K_i^u T_{zc} + K_f^u}{T_s}\rceil,
	\label{eq:normalized arrival time}
\end{equation}
where $\lceil . \rceil$ represents the round-up operation.

In TDD system, UEs estimate uplink signal attenuation based on SSBs and adjust the amplitude of preamble signal using open loop power control.
Then, it is reasonable to assume that the amplitudes of UEs' preambles received by the satellite are approximately equal, which means $A_uP_i^u$ in (\ref{eq:gNBreceived signal}) can be treated as a constant.
In addition, the noise term has no impact on the correlation property of ZC sequence and hence noise term is negligible in the following analysis.
Meanwhile, when accurate frequency pre-compensation operation is performed at UE side, the frequency offset in (\ref{eq:gNBreceived signal}) can be set to zero.

%---------------------------------
\subsection{The general procedure of 2-step TA estimation }\label{sec:TA_estimation}
%---------------------------------

\begin{figure*}[!t]
	\center
	\includegraphics[width=5.5in]{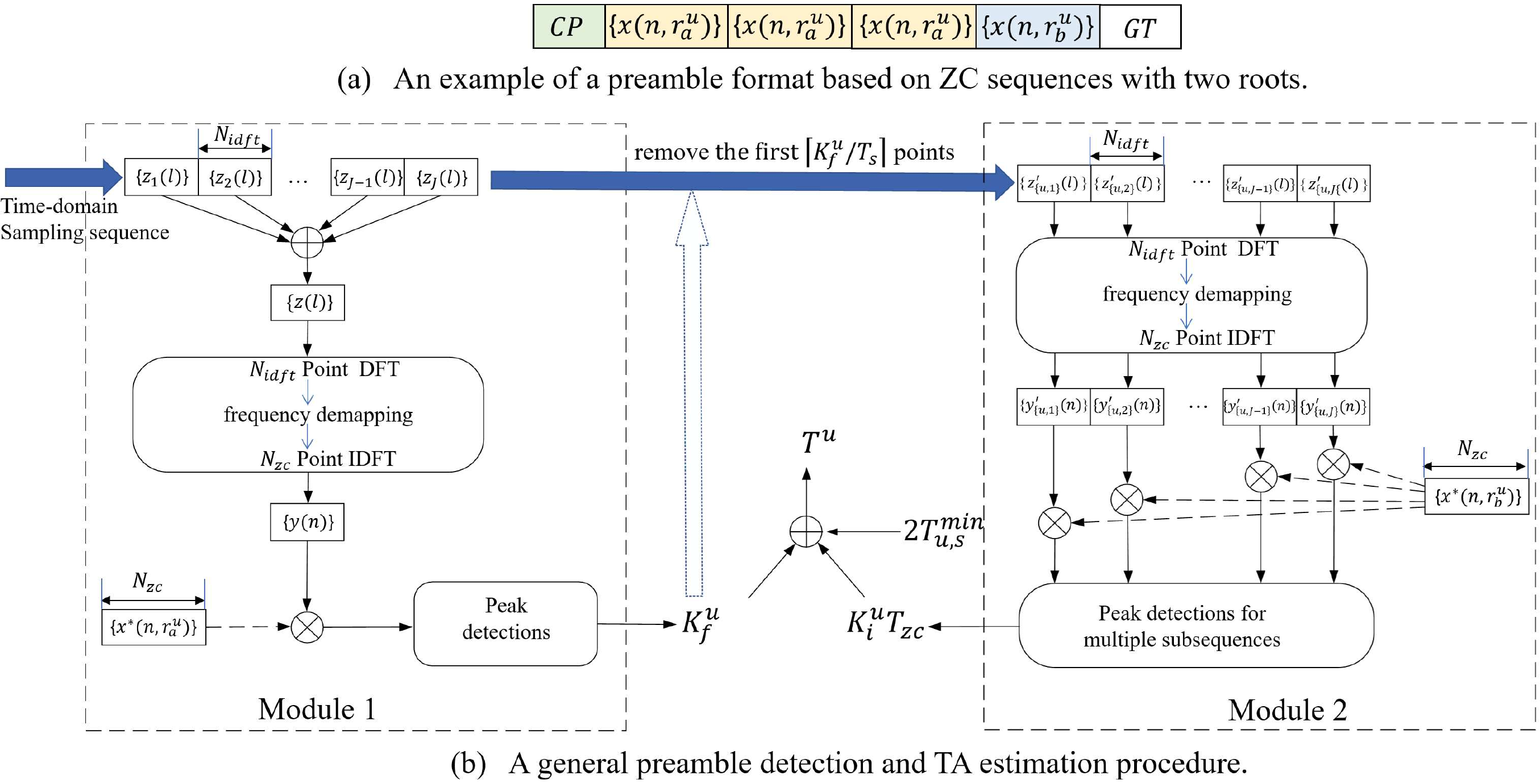}
	\caption{An example preamble format design and a general preamble detection procedure.}\label{fig2}
\end{figure*}

In (\ref{eq:TA repre}), $T^u$ is expressed with $T_{u,s}^{min}$, $K_f^u $ and $K_i^u $, where $T_{u,s}^{min}$ depends on the setting of the satellite and is known to satellite.
Therefore, with the estimated $K_f^u$ and $K_i^u$, the satellite can estimate TA.
As shown in Fig.~\ref{fig2}(b), a general preamble detection procedure consists of two modules at the satellite for $K_f^u $ and $K_i^u $ estimation, respectively \cite{zhenli} \cite{huawei}.
To acquire $K_f^u$, module 1 divides the received time-domain preamble signals from multiple UEs into multiple subsequences, each with a length of $N_{idft}$.
Denote the $j$-th subsequence as $\{z_j(l)\}$, where $j=1,2,...,J$ with $J=Z_l+\lceil T_{gt}/T_{zc}\rceil$.
Here $T_{gt}$ represents the duration of the guard time in preamble format.
By referring to\cite{zhenli}, we can obtain $K_f^u$ based on sequence $\{z(l)\}$ that is the sum of all subsequences $\{z_j(l)\}$, and $z(l)$ is written as
\begin{equation}
	z(l) =\sum_{j} z_j(l) =\sum_{u} \sum_{i} z_i^u(l).
	\label{eq:TA estimation}
\end{equation}
Next, the satellite performs $N_{idft}$ point DFT, frequency demapping and $N_{zc}$ point IDFT on sequence $\{z(l)\}$ to obtain sequence $\{y(n)\}$, and $y(n)$ is given by~\cite{zhenli}
\begin{equation}
	y(n)= \sum_{u}\sum_{i} x(n-\Theta_u,r_i^u),
	\label{eq:received_sum_zc_sequence}
\end{equation}
where
\begin{equation}
	\Theta_u =\lceil \frac {K_f^u N_{zc}}{T_{zc}}\rceil.
	\label{eq:fractional_TA}
\end{equation}

Subsequently, a power delay profile (PDP) computation is performed on $\{y(n)\}$ and a ZC sequence $\{x(n.r_a^u)\}$, which generates sequence $\{PDP(m,r_a^u)\}$ with $m=0,1,...,N_{zc}-1$ as follows
\begin{equation}
	\begin{aligned}
		 PDP(m,r_a^u) &=\left|\sum_{n=0}^{N_{zc}-1}\frac {y(n)x^*(n-m,r_a^u)}{{N_{zc}}}          \right|^2 \\
	     &=\left|\frac {1}{{N_{zc}}} \sum_{u} \sum_{i} C_i^u(m)\right|^2,
	\end{aligned}
	\label{eq:cal frac_pdp}
\end{equation}
where $(.)^*$ represents conjugate operation and
\begin{equation}
	\begin{aligned}
		C_i^u(m)&=\sum_{n=0}^{N_{zc}-1} x(n-\Theta_u,r_i^u)x^*(n-m,r_a^u)\\
		&=\sum_{n=0}^{N_{zc}-1}  \exp[j \frac{\pi r_a^u(n-m)\left(n-m+1 \right)}{N_{zc}}\\
		&- j \frac{\pi r_i^u(n-\Theta_u)\left(n-\Theta_u+1 \right)}{N_{zc}}].
	\end{aligned}
	\label{eq:cor_cal}
\end{equation}
It can be observed that ${C_i^u(m)}$ reaches the maximum value $N_{zc}$, when $r_i^u=r_a^u$ and $m = \Theta_u$.
Therefore, if a UE transmits several ZC sequences with root $r_a^u$, $\{PDP(m,r_a^u)\}$ takes a local maximum value at the point $m^* = \Theta_u$.
Based on the above analysis and equation (\ref{eq:fractional_TA}), module 1 in Fig.~\ref{fig2}(b) obtains $K_f^u$ based on a peak search procedure, and $K_f^u$ is given by
\begin{equation}
	K_f^u =\lceil \frac {m^* T_{zc}}{N_{zc}} \rceil.
	\label{eq:fractional_TA2}
\end{equation}

After estimating $K_f^u$, module 2 further calculates $K_i^u$ based on PDP computations.
In the beginning, module 2 removes the first $\lceil K_f^u/T_s \rceil$ sampling points in the PRACH and then divides the rest sampling point sequence into multiple subsequences $\{z_{u,j}^\prime(l)\}$, each with a length of $N_{idft}$.
Then, module 2 performs $N_{idft}$ point DFT, frequency demapping and $N_{zc}$ point IDFT on $\{z_{u,j}^\prime(l)\}$, and the resulted sequence is denoted as $\{y_{u,j}^\prime(n)\}$, where
\begin{equation}
	y_{u,j}^\prime(n)= x(n,r_{j-K_i^u}^u)+ I_j^u(n).
	\label{eq:y_si}
\end{equation}
In (\ref{eq:y_si}), $I_j^u(n)$ represents the interference from other UEs' preambles and is expressed as
\begin{equation}
	I_j^u(n)= \sum_{u^\prime \in \mathcal{U} \setminus {u}} \sum_{i} a_{u,i}^{u^\prime}(n) x(n+\Theta_u-\Theta_{u^\prime},r_i^{u^\prime}),
	\label{eq:interference_term}
\end{equation}
where $a_{u,i}^{u^\prime}(n)=1$ indicates the transmitted $i$-th ZC sequence from UE $u^\prime$ exists in $I_j^u(n)$ and vice versa.

Later on, module 2 performs PDP computation between sequence $\{y_{u,j}^\prime(n)\}$ and a ZC sequence with root $r_b^u$.
If the resulted PDP sequence $\{PDP(m,r_b^u)\}$ obtains a local maximum value at $m=0$ and $j=j^*$, the sequence $\{y_{u,j}^\prime(n)\}$ contains a ZC sequence with root $r_b^u$.
Then, $K_i^u$ is calculated as
\begin{equation}
	K_i^u= j^*-b^u
	\label{eq:interger_ta},
\end{equation}
where $b^u$ is the position index of ZC sequence with root $r_b^u$ in preamble and $b^u \in \{1,2,...,Z_l\}$.

Note that the above estimation procedure of $K_f^u$ and $K_i^u$ can also be applied to other two preamble format options advised by 3GPP.
The differences are that $b^u=1$ in Option-1 and unscrambling step is additionally required in Option-2.

%---------------------------------
\subsection{Proposed enhanced TA estimation procedure}\label{sec:timing_overview}
%---------------------------------

\begin{figure}[!t]
	\center
	\includegraphics[width=1.8in]{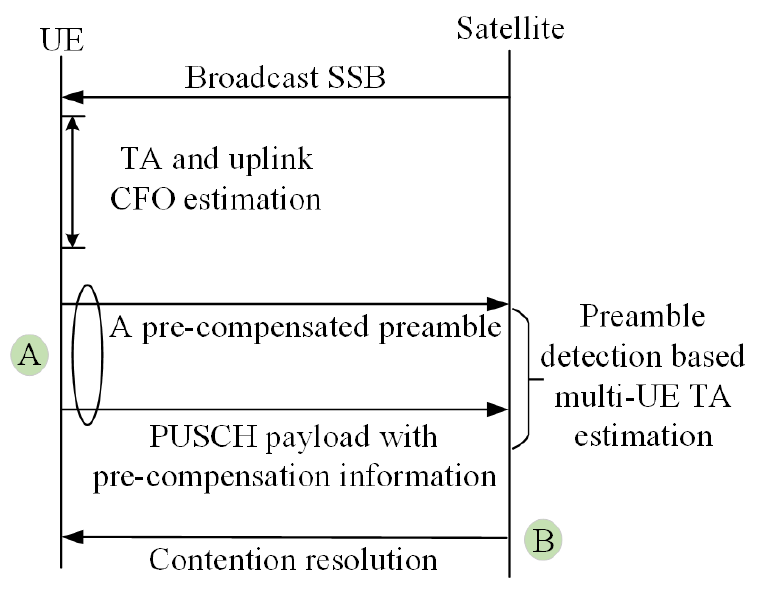}
	\caption{An enhanced TA estimation procedure in 2-step random access.}\label{fig3}
\end{figure}

In this subsection, an enhanced TA estimation procedure for 5G LEO networks is proposed to tackle the challenges arising from large CFO and differential TA.
As shown in Fig.~\ref{fig3}, our proposal consists of a time-frequency pre-compensation estimation step at the UE side, preamble generation and transmission step, a multi-UE preamble detection and TA estimation step at the satellite, and a contention resolution step.
The reason for time-frequency pre-compensation at UE side is that it can improve the time-domain resource utilization and meanwhile post-compensation at the satellite might cause intolerable computational complexity due to multi-UE access.
In terms of preamble format design, our proposal removes CP and extends the length of guard time as well as preamble compared with that in terrestrial networks.
Moreover, our proposal adopts a differential power allocation strategy and flexible cascading order of ZC sequences with different roots.

After involving time pre-compensation, actual TA $T^u$ between UE $u$ and the target satellite can be rewritten as
\begin{equation}
	T^u=T_{pre}^u +K_i^u T_{zc} + K_f^u,
	\label{eq:acutal_TA repre}
\end{equation}
where $T_{pre}^u$ represents the time pre-compensation value of UE $u$.
Via preamble detection, the satellite can obtain the TA pre-compensation error of UE $u$, which is the sum of the last two terms in equation (\ref{eq:acutal_TA repre}).
In the contention resolution step, UEs obtain TA pre-compensation errors that guide them to achieve uplink synchronization, while the satellite acquires the full TA of UE $u$ according to $T_{pre}^u$ transmitted in physical uplink shared channel (PUSCH) payload.
More details about pre-compensation at UE side and preamble format design method will be presented in Section \ref{sec:time-frequency pre-compensation} and Section \ref{sec:proposed_preamble_design2}, respectively.
Note that the above process can also be extended to 4-step random access procedure in \cite{38.821}, and the time-frequency pre-compensation only relies on SSB measurement, which does not cause additional signalling overhead before random access.

%---------------------------------
\section{Time-frequency Pre-compensation Method With Downlink Synchronization Signals }\label{sec:time-frequency pre-compensation}
%---------------------------------

In this section, we investigate time-frequency pre-compensation method in 5G LEO networks based on SSBs, aiming at reducing the uncertainty of preamble arrival time and frequency offset observed at the satellite.

%---------------------------------
\subsection{Problem formulation}
%---------------------------------
Owning to recent research developments in SSB detection in 5G LEO networks, downlink synchronization can reach a near-optimal performance\cite{f_est1,f_est2,f_est3}.
With ephemeris data provided by downlinks, each UE has the ability to infer the position and velocity of multiple satellites in the initial cell search phase.
Define $\mathcal{S}=\{1,2,...,S\}$ as the set of satellites detected by UE $u$.
As shown in Fig.~\ref{fig1}, the position of satellite $s$, which operates at carrier frequency $f_s$, is characterized by  $(x_s,y_s,z_s)$ in Earth Centered Earth Fixed (ECEF) coordinates.
Define $(x_u,y_u,z_u)$ and $(v_s^x,v_s^y,v_s^z)$ as the position of a geostationary UE $u$ and the velocity of satellite $s$ in ECEF, respectively.
The doppler shift $f_{s}^u$ between UE $u$ and satellite $s$ is calculated as
\begin{equation}
f_{s}^u=f_s \frac{\left[v_s^x \left(x_s-x_u\right)+ v_s^y \left(y_s-y_u\right)+ v_s^z \left(z_s-z_u\right)\right]}{c\sqrt{\left(x_s-x_u\right)^2+\left(y_s-y_u\right)^2+\left(z_s-z_u\right)^2}}
\label{eq:doppler spectrum}
\end{equation}

Assuming that UE $u$ is on the surface of the Earth treated as a regular sphere.
The position of UE $u$ is estimated based on spherical coordinate system instead of Cartesian coordinate system.
The major advantage of spherical coordinate system is that it can reduce the number of parameters needed to estimate.
In spherical coordinate system, the position of UE $u$ is characterized by $\left(R, \theta_u, \varphi_u \right)$ with $R$, $\theta_u$, $\varphi_u$ representing the radius of the Earth, the polar angle and the azimuth angle, respectively.
Then, we have
\begin{equation}
\label{eq:xyz}
\left\{
             \begin{array}{lr}
             x_u=R\cos(\theta_u)\cos(\varphi_u),  & \\
             y_u=R\sin(\theta_u)\cos(\varphi_u),  & \\
             z_u=R\sin(\varphi_u).
             \end{array}
\right.
\end{equation}

The orbital altitude of satellites is denoted as $h$.
According to (\ref{eq:doppler spectrum}), the doppler shift $f_{s}^u$ can be rewritten as
\begin{equation}
\begin{aligned}
f_{s}^u =f_s (\frac{ v_s^x(x_s-x_u)+v_s^y(y_s-y_u)+v_s^z(z_s-z_u)  }{c\sqrt{(R+h)^2+R^2-2x_sx_u-2y_sy_u-2z_sz_u}}).
\end{aligned}
\label{eq:do1}
\end{equation}

Define two auxiliary variables $\gamma_{s,u}^1$ and $\gamma_{s,u}^2$ as follow
\begin{equation}
\left\{
             \begin{array}{lr}
             \gamma_{s,u}^1= v_s^x(x_{s} -x_u )+v_s^y(y_{s} - y_u)+v_s^z(z_{s} - z_u),  &\\
             \gamma_{s,u}^2 = \sqrt {(R+h)^2+R^2 -2x_{s}x_u-2y_{s}y_u-2z_{s}z_u}.
             \end{array}
\right.
\end{equation}
Note that $(x_u,y_u,z_u)$ in (20) can be calculated through $\theta_u$ and  $\varphi_u$ in (\ref{eq:xyz}).
Then, doppler shift in (\ref{eq:do1}) can be expressed as a function about $\theta_u$ and  $\varphi_u$, which is given by
\begin{equation}
\begin{aligned}
f_{s}^u(\theta_u, \varphi_u) =\frac{ f_s \gamma_{s,u}^1 }{c\gamma_{s,u}^2}.
\end{aligned}
\end{equation}

Define $f_{lo}^u$ as the frequency offset between UE $u$ and satellites caused by local oscillators.
Then, the downlink CFO denoted as $f_{u,s}^{down}$ is given by
\begin{equation}
f_{u,s}^{down}=f_{lo}^u+f_{s}^u.
\label{eq:frequency_cal}
\end{equation}
Since TA is twice the propagation delay, we can calculate it based on the position of UE $u$ and satellite $s$, which is
\begin{equation}
T^u=\frac{2\sqrt{\left(x_s-x_u\right)^2+\left(y_s-y_u\right)^2+\left(z_s-z_u\right)^2}}{c}.
\label{eq:TA_cal}
\end{equation}

Since both TA and frequency offset relate to UE position, TA and uplink CFO estimations are converted into a UE localization problem.
In traditional satellite networks, the positioning methods can be summarized into two categories\cite{position_error}, namely code-based positioning and doppler-based positioning.
Owning to the fast movement of LEO satellites, a large amount of downlink doppler shift observations in a short period can be acquired by a UE.
Hence, we introduce a method to estimate UE's location based on doppler shift in the following.
The doppler-based UE positioning problem is formulated as
\begin{align}
&\min\limits_{\hat \theta_u, \hat \varphi_u, \hat f_{lo}^u}  \quad \sum_{s} \left( \hat f_{u,s}^{down}- f_{u,s}^{down} \right)^2
& \begin{array}{r@{\quad}l@{}l@{\quad}l}
\end{array}
\label{problem}
\end{align}
\begin{equation}
s.t. \;\;\; \;\;\hat f_{u,s}^{down}= \hat f_{s}^u+\hat f_{lo}^u+ \sigma_s^u ,
\label{eq:problem_constrain}
\end{equation}
where $\hat f_{u,s}^{down}$ is the estimated downlink frequency offset, $f_{u,s}^{down}$ is downlink frequency offset output by downlink synchronization algorithms,  $\hat f_{s}^u$  and $\hat f_{lo}^u$ represent the estimated doppler shift and frequency offset caused by local oscillators under the estimated position $\left(R, \hat \theta_u, \hat \varphi_u \right)$, respectively. $\sigma_s^u$ is the downlink frequency offset measurement error.

%---------------------------------
\subsection{Location-based TA and CFO estimation}
%---------------------------------
In this part, we estimate TA and uplink CFO based on an iterative least-square algorithm.
At first, the algorithm obtains an estimated position of UE and frequency offset caused by local oscillators by solving the optimization problem (24).
Then, UE $u$ estimates uplink CFO and TA between it and the target satellite.

Denote estimation vector of downlink frequency offset of UE $u$ as $\hat {\boldsymbol f}^{down} = ( \hat f_{u,1}^{down}, \hat f_{u,2}^{down},...,\hat f_{u,S}^{down})$, where $S$ represent the number of detected SSB from different satellites.
In problem (\ref{problem}), $\hat {\boldsymbol f}^{down}$ is a function of variable vector $(\hat \theta_u, \hat \varphi_u, \hat f_{lo}^u)$.
Denote $\boldsymbol J$ as the Jacobian matrix of  $\hat {\boldsymbol f}^{down}$, which is given by
\begin{equation}\label{J}
\boldsymbol J=
\begin{bmatrix}
\frac{\partial \hat f_{u,1}^{down}}{\partial \theta} & \frac{\partial \hat f_{u,1}^{down}}{\partial  \varphi }  & \frac{\partial \hat f_{u,1}^{down}}{\partial f_{lo}^u}   \\
\frac{\partial \hat f_{u,2}^{down}}{\partial \theta} & \frac{\partial \hat f_{u,2}^{down}}{\partial  \varphi } & \frac{\partial \hat f_{u,2}^{down}}{\partial f_{lo}^u}   \\
\cdots & \cdots  & \cdots   \\
\frac{\partial \hat f_{u,S}^{down}}{\partial \theta} & \frac{\partial \hat f_{u,S}^{down}}{\partial  \varphi } & \frac{\partial \hat f_{u,S}^{down}}{\partial f_{lo}^u}
\end{bmatrix},
\end{equation}
where $\frac{\partial \hat f_{u,s}^{down}}{\partial \theta}=\frac{\partial \hat f_{s}^u}{\partial \theta}$, $\frac{\partial \hat f_{u,s}^{down}}{\partial \varphi}=\frac{\partial \hat f_{s}^u}{\partial \varphi}$ since variables $\hat f_{lo}^u$ and $\sigma_s^u$ are independent of $\theta$ and $\varphi$,
and $\frac{\partial \hat f_{u,s}^{down}}{\partial f_{lo}^u} =1$ since variables $\hat f_{s}^u$ and $\sigma_s^u$ are independent of $f_{lo}^u$.
Meanwhile, according to equation (18)-(21), we have
\begin{equation}
\begin{aligned}
\frac{\partial \hat f_{s}^u}{\partial \theta}= \frac{f_s \left(v_s^x \hat y_u - v_s^y \hat x_u\right)}{c \hat \gamma_{s,u}^2}- \frac{f_s (x_{s}\hat y_u -y_{s}\hat x_u )\hat \gamma_{s,u}^1}{c (\hat \gamma_{s,u}^2)^3}.
\end{aligned}
\end{equation}
Since $\hat x_u$ and $\hat y_u$ are the function of $\theta_u$ and $\phi_u$ according to (\ref{eq:xyz}),
to simplify the representation of $\frac{\partial \hat f_{s}^u}{\partial \varphi}$, define auxiliary variables $\alpha_u^1$, $\alpha_u^2$ and $\alpha_u^3$ as follows:
\begin{equation}
\left\{
             \begin{array}{lr}
             \alpha_u^1= -\frac{\partial x_u}{\partial \varphi}  =R\cos(\theta_u)\sin(\varphi_u),  &\\
             \alpha_u^2= -\frac{\partial y_u}{\partial \varphi} =R\sin(\theta_u)\sin(\varphi_u),  &\\
             \alpha_u^3=\frac{\partial z_u}{\partial \varphi}=R\cos(\varphi_u).
             \end{array}
\right.
\end{equation}
Then, based on (\ref{eq:xyz}) and (28), $\frac{\partial \hat f_{s}^u}{\partial \varphi}$ can be expressed as
\begin{equation}
\begin{aligned}
\frac{\partial \hat f_{s}^u}{\partial \varphi} &=\frac{f_s \left(v_s^x \hat \alpha_u^1+ v_s^y \hat \alpha_u^2 - v_s^z \hat \alpha_u^3 \right)}{c \hat \gamma_{s,u}^2}\\
& - \frac{f_s (x_{s}\hat \alpha_u^1 +y_{s}\hat \alpha_u^2 -z_{s}\hat \alpha_u^3)\hat \gamma_{s,u}^1}{c (\hat \gamma_{s,u}^2)^3},
\end{aligned}
\end{equation}
where $\hat x_u,\hat y_u,\hat z_u,\hat \gamma_{s,u}^1, \hat \gamma_{s,u}^2, \hat \alpha_u^1, \hat \alpha_u^2, \hat \alpha_u^3$ are calculated based on the estimated positioning $(R,\hat \theta_u, \hat \varphi_u)$.

Denote $ \boldsymbol {\hat\mu}_u=[\hat \theta_u, \hat \varphi_u, \hat f_{lo}^u]$ as the estimated result of problem (\ref{problem}).
Note that the expression of $\hat f_{u,s}^{down}$ in equation (\ref{eq:problem_constrain}) is not linear, which means that the least-square algorithm cannot be applied directly.
Define the vector of downlink frequency offset $f_{u,s}^{down}$ output by downlink synchronization algorithms as $\boldsymbol f^{down}=(  f_{u,1}^{down}, f_{u,2}^{down},..., f_{u,S}^{down})$.
Here, we perform a linear approximation of $\boldsymbol f^{down}$ based on Jacobian matrix $\boldsymbol J$ and ignore higher-order terms as well as estimation error, which is given by
\begin{equation}
\begin{aligned}
\boldsymbol f^{down}= \hat {\boldsymbol f}^{down}(\boldsymbol {\hat\mu_u}) +\boldsymbol J (\boldsymbol \mu_u-\boldsymbol {\hat\mu_u})
\end{aligned}
\label{eq:linear}
\end{equation}
where frequency offset vector $\hat {\boldsymbol f}^{down}(\boldsymbol {\hat\mu_u})$ is calculated by equation (\ref{eq:problem_constrain}) based on the current estimation result $\boldsymbol {\hat\mu}_u$, $\boldsymbol \mu_u$ is the actual result of problem (\ref{problem}).
After reorganizing equation (\ref{eq:linear}), we obtain
\begin{equation}
 \Delta \boldsymbol { f}=\boldsymbol J(\boldsymbol \mu_u-\boldsymbol {\hat\mu}_u),
\label{eq:update_location}
\end{equation}
where $\Delta \boldsymbol { f}={\boldsymbol f}^{down}-\hat {\boldsymbol f}^{down}(\boldsymbol {\hat\mu_u})$.
By following the principle of least-square algorithm, a more accurate result $\boldsymbol {\hat \mu}_u^\prime$ of problem (\ref{problem}) is obtained by~\cite{Least-Squares}
\begin{equation}
 \boldsymbol {\hat \mu}_u^\prime = \hat {\boldsymbol \mu}_u + \Delta \boldsymbol { \mu}_u,
\end{equation}
where
\begin{equation}
  \Delta \boldsymbol { \mu}_u = (\boldsymbol J^T \boldsymbol J)^{-1} \boldsymbol J^T \Delta \boldsymbol f.
\end{equation}
Finally, we can obtain more accurate position estimation results by repeating the above steps.

\textcolor{yellow} {
\begin{algorithm}[htbp]
	\begin{algorithmic}[1]
        \scriptsize
		\caption{TA and uplink CFO estimation algorithm based on iterative least squares method}
		\label{alg:1}
		\State Initiate $\boldsymbol {\hat \mu}_u$ with $\hat f_{lo}^u=0$;
		\For{$k=1:K$}
		\State Obtain the  Jacobian matrix $\boldsymbol J$ according to (\ref{J});
		\State Calculate  $\Delta \boldsymbol { f}$ according to (\ref{eq:update_location});
		\State Obtain $\boldsymbol {\hat \mu}_u^\prime$ with $\boldsymbol {\hat \mu}_u^\prime =  \hat {\boldsymbol \mu}_u + \Delta \boldsymbol { \mu}_u$;
        \If{$ \Vert \Delta \boldsymbol {\mu}_u \Vert \leq \sigma  $}
        \State break;
        \EndIf
        \State Update $\boldsymbol {\hat\mu}_u$ with $\boldsymbol {\hat\mu}_u=\boldsymbol {\hat \mu}_u^\prime$;
        \EndFor
        \State According to (\ref{eq:TA_cal}) and (\ref{eq:f_e}), estimate TA and uplink CFO based on $\boldsymbol {\hat\mu}_u^\prime$;
		\State Output $\hat T^u$ and $\hat f_{u,s}^{up}$.
	\end{algorithmic}
\end{algorithm}
}

\begin{figure}[!t]
	\center
	\includegraphics[width=2.1in]{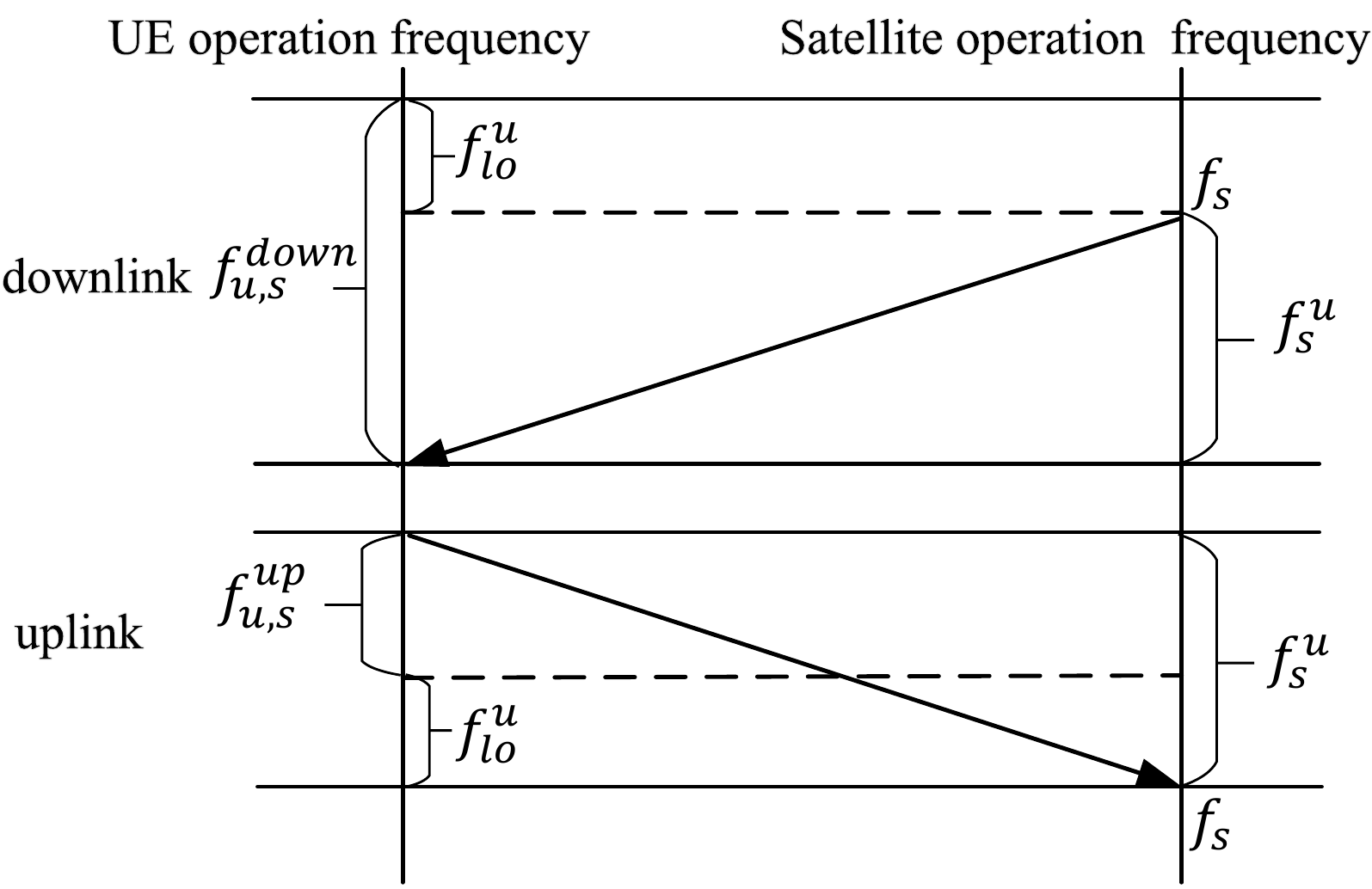}
	\caption{Uplink frequency offset pre-compensation considering the impact of local oscillators.}\label{fig4}
\end{figure}

The location-based TA and CFO estimation process is shown in Algorithm~\ref{alg:1}.
At the beginning,
$\hat f_{lo}^u$ is set to zero.
When reaching the maximum number of iterations or the update step size is less than the preset threshold $\sigma $, Algorithm 1 stops updating $ {\boldsymbol {\hat\mu_u}}$.
Finally, Algorithm 1 calculates TA and uplink CFO based on the estimated UE location and $\hat f_{lo}^u$, and then they are taken as the uplink TA and frequency pre-compensation value at UE side.
Particularly, as illustrated in Fig.~\ref{fig4}, estimated uplink CFO $\hat f_{u,s}^{up}$ is obtained based on the downlink CFO and estimated $\hat f_{lo}^u$, which is given by
\begin{equation}
\hat f_{u,s}^{up}  =\hat f_{s}^u-\hat f_{lo}^u=f_{u,s}^{down} -2\hat f_{lo}^u.
\label{eq:f_e}
\end{equation}

%---------------------------------
\subsection{A discussion about the performance of Algorithm 1}
%---------------------------------

In this subsection, we discuss the impact of positioning error $\left( \Delta x,\Delta y,\Delta z \right)$ in Cartesian coordinate system on the accuracy of the estimated  $\hat T^u$ and $\hat f_{u,s}^{up}$.
Denote the difference between $\hat T^u$ and actual TA $T^u$ as TA estimation error $T_e^u$, and denote the difference between $\hat f_{u,s}^{up}$ and actual uplink CFO $f_{u,s}^{up}$ as uplink CFO estimation error $f_e^u$.
In (\ref{eq:TA_cal}), it can be observed that an accurate positioning result contributes to a smaller $T_e^u$ and hence improving time pre-compensation accuracy.
For the impact of positioning on $\hat f_{u,s}^{up}$, define $\Delta f_{s}^u$  and $\Delta f_{lo}^u$ as the estimation errors of doppler shift and frequency offset introduced by local oscillators, respectively.
According to (\ref{eq:f_e}), the relationship between $f_e^u$ and $\Delta f_{lo}^u$ is expressed as
\begin{equation}
\begin{aligned}
f_e^u & = \hat f_{u,s}^{up}- f_{u,s}^{up}\\
&=f_{u,s}^{down} -2\hat f_{lo}^u -(f_{u,s}^{down} -2 f_{lo}^u) =2\Delta f_{lo}^u.
\end{aligned}
\label{eq:relationship_f}
\end{equation}
Therefore, $f_e^u$ is proportional to $\Delta f_{lo}^u$.
Hence, we only need to analyze the relationship between positioning error $\left( \Delta x,\Delta y,\Delta z \right)$ and $\Delta f_{lo}^u$.
Based on equations (\ref{eq:doppler spectrum}), (\ref{eq:do1}) (20) and (23), the actual doppler shift and estimated doppler shift can be written as
\begin{equation}
\begin{aligned}
f_s^u &= f_s(\frac{ v_s^x(x_s-x_u)+v_s^y(y_s-y_u)+v_s^z(z_s-z_u)  }{c\sqrt{(x_u-x_s)^2+(y_u-y_s)^2+(z_u-z_s)^2}})\\
&=\frac{f_s\gamma_{s,u}^1}{c(\frac{cT^u}{2})},
\end{aligned}
\end{equation}
and
\begin{equation}
\begin{aligned}
\hat f_s^u &= f_s(\frac{\gamma_{s,u}^1- v_s^x\Delta x-v_s^y\Delta y-v_s^z\Delta z)  }{c(\frac{c(T^u+T_e^u)}{2})},
\end{aligned}
\end{equation}
Hence, $\Delta f_{s}^u$ can be written as
\begin{equation}
\begin{aligned}
\Delta f_{s}^u&=f_s^u-\hat f_s^u
\\&= \frac{2f_s}{c^2}(\frac{\gamma_{s,u}^1}{T^u}-\frac{\gamma_{s,u}^1- v_s^x\Delta x-v_s^y\Delta y-v_s^z\Delta z}{T^u+T_e^u}))\\
&=\frac{f_s(v_s^x\Delta x+v_s^y\Delta y+v_s^z\Delta z))}{c(\frac{cT^u}{2}+\frac{cT_e^u}{2})} + \frac{\frac{f_s\gamma_{s,u}^1}{c(\frac{cT^u}{2})}}{1+\frac{T^u}{T_e^u}}\\
&=\frac{f_s \left[v_s^x\Delta x +  v_s^y \Delta y +v_s^z \triangle z \right]}{c \left[d_{u,s}+ \frac{c  T_e^u}{2} \right]}
+ \frac{f_{s}^u}{1+\frac{2d_{u,s}}{c T_e^u}},
\end{aligned}
\label{eq:delta_f}
\end{equation}
where $d_{u,s}$ is the distance between UE $u$ and satellite $s$.

According to (\ref{problem}),  $\Delta f_{lo}^u$ can be written as
\begin{equation}
\Delta f_{lo}^u=  \sum_{s=1}^S \frac{1}{S} \left( \Delta f_{s}^u+ \sigma_s^u \right).
\label{eq:delta_flo}
\end{equation}
Considering that one-way propagation delay in LEO satellite networks is much larger than $T_e^u$,
$\frac{c  T_e^u}{2}$ can be ignored compared to $d_{u,s}$.
Therefore, with the positioning accuracy improving, the first term on the right side of (\ref{eq:delta_f}) decreases and meanwhile the second term also decreases, which
contributes to smaller $\Delta f_{lo}^u$ and hence $f_e^u$.
Note that large downlink frequency measurement errors $\sigma_s^u$ can cause a considerable $\Delta f_{lo}^u$ and a larger $f_e^u$.
To overcome this issue, the precision of Algorithm 1 can be improved by more CFO measurements\cite{position_error}.

%---------------------------------
\section{Inter-preamble Interference Analysis and Enhanced Preamble Format Design}\label{sec:preamble_design1}
%---------------------------------
After introducing the method for time-frequency pre-compensation step in Fig.~\ref{fig3}, in this section, we further present our enhanced preamble design.
Specifically, the upper bound of inter-preamble interference is derived at first. Based on the result, a flexible preamble format is designed and the advantages of our proposal are highlighted.
At last, we provide two ways of implementation in 5G LEO networks.

%---------------------------------
\subsection{The upper bound of inter-preamble interference}\label{sec:interference_analysis}
%---------------------------------

Firstly, we provide several definitions as follows.
\begin{itemize}
  \item Full-period cross-correlation: A correlation operation between two complete ZC sequences with different roots.
  \item Partial-period cross-correlation: A correlation operation between a complete ZC sequence and an incomplete ZC sequence, and they have different roots.
\end{itemize}
Then, we start investigating inter-preamble interference among UEs based on preamble format Option-3 introduced in Section~\ref{sec:preamble_option}.

Firstly, the autocorrelation peak ratio of module 1 to module 2 is analyzed.
As shown in Fig.~\ref{fig2}(b), all ZC sequences with root $r_a^u$ in a preamble are accumulated to obtain $K_f^u$, while $K_i^u$ is acquired only based on one ZC sequence with root $r_b^u$, which causes such correlation peak ratio to be $(Z_l-1)^2$, where $Z_l$ is the number of ZC sequences in a preamble.
Considering that a lower PDP peak is more sensitive to interference in the TA estimation procedure relying on peak detection, the estimation of $K_i^u$ significantly limits the whole performance\cite{huawei}.

\begin{figure}[!t]
	\center
	\includegraphics[width=2.8in]{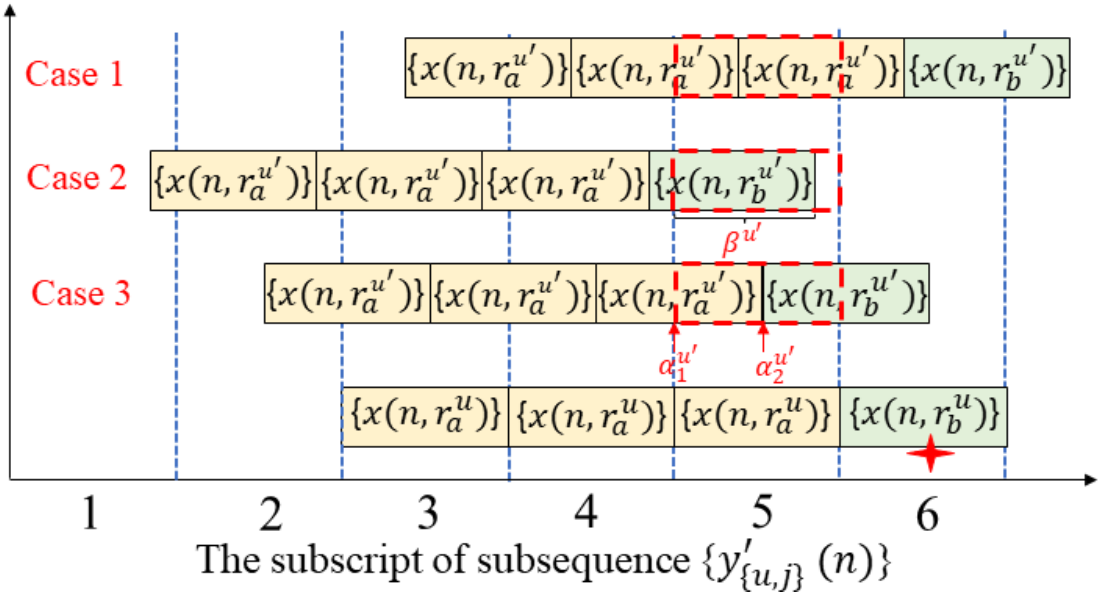}
	\caption{ An estimation example of $K_i^u$ for UE $u$ with the correct $K_i^u$ being 2.
 In this example, the preamble sequences of UE $u^\prime$ bring large inter-preamble interference.
 }\label{fig5}
\end{figure}

Fig.~\ref{fig5} illustrates all the cases of inter-preamble interference between UE $u$ and $u^\prime$ in module 2,
in which the satellite intends to estimate $K_i^u$ for UE $u$ and the ZC sequences sent by UE $u^\prime$ are interfering sources.
Since satellites perform linear shift operation on the received preamble sampling sequence according to the estimated $K_f^u$ in module 1, subsequences $\{y_{u,j}^\prime(n)\}$ consist of complete ZC sequences from UE $u$ and other incomplete ZC sequences from UE $u^\prime$.
Considering that $K_i^u$ is uncertain before the preamble detection procedure, satellites must calculate PDP computations between each potential sequence $\{y_{u,j}^\prime(n)\}$ and a ZC sequence $\{x(n,r_b^u)\}$ in module 2 and then perform peak search procedure.
If no UE interferes with UE $u$, satellites will observe only a peak occurring in the PDP computation result of subsequence $\{y_{u,6}^\prime(n)\}$, which is indicated by the red star in Fig.~\ref{fig5}.
In this case, $K_i^u=6-4=2$.
However, for a detected subsequence $\{y_{u,5}^\prime(n)\}$ calculated by (\ref{eq:y_si}) in Fig.~\ref{fig5}, interference term in it given by (\ref{eq:interference_term}) contains a complete ZC sequence $\{x(n,r_a^{u^\prime})\}$ in Case 1,  an incomplete ZC sequence $\{x(n,r_b^{u^\prime})\}$ in Case 2 and two incomplete ZC sequences $\{x(n,r_a^{u^\prime})\}$ and $\{x(n,r_b^{u^\prime})\}$ in Case 3.
In addition, in TA estimation procedure, satellites can first estimate the noise strength in PDP results and then carry out peak detection.
For example, the values of the low amplitude points in PDP results are summed and averaged, then the averaged result is treated as noise and filtered out.
The interference in Case 1 caused by a complete ZC sequence $\{x(n,r_a^{u^\prime})\}$ can be considered as a constant $\frac{1}{\sqrt{N_{zc}}}$\cite{Partial-Period}, which means that interference in Case 1 can be eliminated as a part of the noise in the noise filtering process.
Therefore, we mainly focus on Case 2 and Case 3.

\subsubsection{The upper bound of interference in Case 2}\label{sec:case2_analysis}
\
\newline
\indent
We first derive the upper bound of interference in Case 2 and further extend the result to Case 3.
Denote the amplitude ratio of the interfering sequence to a ZC sequence $\{x(n,r_b^u)\}$ generated by satellites as $k$, and $k=1$ means that both sequences have equal amplitude.
Define two non-negative integers $\alpha^{u^\prime}$ and $\beta^{u^\prime}$ with $\alpha^{u^\prime}+\beta^{u^\prime} \leq N_{zc}-1$ as the start point and the length of the interfering sequence from UE $u^\prime$, which can be regarded as a subsequence of a ZC sequence denoted as $\{x(n,r_i^{u^\prime})\}$ in (\ref{eq:zc_expression}).
A windowing sequence $\{\rho_{\alpha^{u^\prime},\beta^{u^\prime}}(n)\}$ is introduced to represent the overlapping area of the interfering sequence and $\{x(n,r_i^{u})\}$, which is given by\cite{windowing_sequence}
\begin{equation}\label{rr1}
\rho_{\alpha^{u^\prime},\beta^{u^\prime}}(n)=\left\{\begin{array}{ll}
1,\; &\alpha^{u^\prime} \leq n\leq \alpha^{u^\prime}+\beta^{u^\prime}, \\
0,\; &otherwise .\\
\end{array}\right.
\end{equation}

Therefore, based on equation (\ref{eq:interference_term}), the incomplete ZC sequence can be represent as $\{x^\prime(n,r_i^{u^\prime})\}$, and $x^\prime(n,r_i^{u^\prime})=kx(n+\Theta_u-\Theta_{u^\prime},r_i^{u^\prime})\rho_{\alpha^{u^\prime},\beta^{u^\prime}}(n)$, where $\Theta_u$ and $\Theta_{u^\prime}$ are defined in equation (\ref{eq:fractional_TA}).
Then, the PDP computation between an incomplete ZC sequence $\{x^\prime(n,r_i^{u^\prime})\}$ and a ZC sequence $\{x(n,r_b^u)\}$ is calculated by
\begin{equation}
\begin{aligned}
&PDP(m,r_b^u)\\
&= \left|\sum_{n=0}^{N_{zc}-1}\frac { \rho_{\alpha^{u^\prime},\beta^{u^\prime}}(n)kx(n+\Theta_u-\Theta_{u^\prime},r_i^{u^\prime}) x^*(n-m,r_b^u)}{N_{zc}}          \right|^2 \\
& \leq \vert \sum_{n=0}^{N_{zc}-1} \frac{k\rho_{\alpha^{u^\prime},\beta^{u^\prime}}(n)}{N_{zc}} \vert^2 \leq (\frac {k\beta^{u^\prime}}{{N_{zc}}})^2.
\end{aligned}
\label{eq:upper_bound1}
\end{equation}
Define $\Upsilon_{r_i^{u^\prime},r_b^u}$ as the result of cross-correlation calculation between sequences $\{x(n+\Theta_u-\Theta_{u^\prime},r_i^{u^\prime})\}$ and $\{x(n,r_b^u)\}$, which is given by
\begin{equation}
\begin{aligned}
\Upsilon_{r_i^{u^\prime},r_b^u}(m)=\sum_{n=0}^{N_{zc}-1} x(n+\Theta_u-\Theta_{u^\prime},r_i^{u^\prime})x^*(n-m,r_b^u).
\end{aligned}
\end{equation}
According to \cite{windowing_sequence}, $PDP(m,r_b^u)$ can also be written as
\begin{equation}
\begin{aligned}
PDP(m,r_b^u)
&=\left|\sum_{n=0}^{N_{zc}-1}\frac {k\widehat \rho_{\alpha^{u^\prime},\beta^{u^\prime}}(n)\Upsilon_{r_i^{u^\prime},r_b^u}(0)}{{N_{zc}}^2}          \right|^2  \\
 &\leq \left [ \sum_{n=0}^{N_{zc}-1}\frac {k\vert \widehat \rho_{\alpha^{u^\prime},\beta^{u^\prime}}(n) \vert \vert \Upsilon_{r_i^u,r_b^u}(0)\vert}{{N_{zc}}^2} \right]^2 \\
&\leq \left [ \sum_{n=0}^{N_{zc}-1}\frac {k\vert \widehat \rho_{\alpha^{u^\prime},\beta^{u^\prime}}(n) \vert }{\sqrt{N_{zc}}^3} \right]^2,
\end{aligned}
\label{eq:32}
\end{equation}
where $\{\widehat \rho_{\alpha^{u^\prime},\beta^{u^\prime}}(n)\}$ is the frequency response of $\{\rho_{\alpha^{u^\prime},\beta^{u^\prime}}(n)\}$, $\vert \Upsilon_{r_i^u,r_b^u}(0)\vert =\sqrt{N_{zc}}$.
The inequality in (\ref{eq:32}) provides another upper bound of $PDP(m,r_b^u)$ based on the property of the absolute value inequality.

As shown in Fig.~\ref{fig5}, we have either $\alpha^{u^\prime}=0$ or $\alpha^{u^\prime}=N_{zc}-1-\beta^{u^\prime}$.
Since the cyclic shift operation of sequence $\{\rho_{\alpha^{u^\prime},\beta^{u^\prime}}(n)\}$ in time domain does not change the amplitude of $\widehat \rho_{\alpha^{u^\prime},\beta^{u^\prime}}(n)$ in frequency domain, $\vert \widehat \rho_{0,\beta^{u^\prime}}(n)\vert =\vert \widehat \rho_{N_{zc}-1-\beta^{u^\prime},N_{zc}-1}(n) \vert$ for any integer $0\leq \beta^{u^\prime} \leq N_{zc}-1$.
Then, $\vert \widehat \rho_{\alpha^{u^\prime},\beta^{u^\prime}}(n) \vert$ in (\ref{eq:32}) can be rewritten as $\vert \widehat \rho_{0,\beta^{u^\prime}}(n)\vert $, which is \cite{Theory}
\begin{equation}
\begin{aligned}
\vert \widehat \rho_{0,\beta^{u^\prime}}(n)\vert = \frac{\vert \sin(\frac{\pi n \beta^{u^\prime} }{N_{zc}}) \vert}{\sin(\frac{\pi n}{N_{zc}})}.
\end{aligned}
\label{eq:PDP5}
\end{equation}

According to\cite{38.221}, the optional values of $N_{zc}$  include 139, 571, 839 and 1151.
Given that $N_{zc}$ is a prime number, based on (\ref{eq:32}) and (\ref{eq:PDP5}), the upper bound of $PDP(m,r_b^u)$ can be further derived as
\begin{equation}
\begin{aligned}
PDP(m,r_b^u) & \leq \left [ \sum_{n=0}^{N_{zc}-1}\frac {k\vert \widehat \rho_{0,\beta^{u^\prime}}(n) \vert}{\sqrt{N_{zc}}^3} \right]^2 \\
& = \frac{k^2}{{N_{zc}}^3} \left [\beta^{u^\prime}+2 \sum_{n=1}^{\frac{N_{zc}-1}{2}} \frac{\vert \sin(\frac{\pi n \beta^{u^\prime}}{N_{zc}}) \vert}{\sin(\frac{\pi n}{N_{zc}})} \right]^2 \\
& < \frac{k^2}{{N_{zc}}^3} \left [\beta^{u^\prime}+2 \sum_{n=1}^{\frac{N_{zc}-1}{2}} \frac{1}{\tan(\frac{\pi n}{N_{zc}})} \right]^2,
\end{aligned}
\label{eq:upper_bound2}
\end{equation}
where the last inequality holds based on the fact that both sequence $\left \{ \vert \sin(\frac{\pi n \beta^{u^\prime} }{N_{zc}}) \vert \right\} $ and $\left \{ \cos(\frac{\pi n}{N_{zc}}) \right\}$ have the same set of elements, when $1\leq n \leq \frac{N_{zc}-1}{2}$.
Since $\tan \left( x \right)$ is a convex function in $\left( 0,\frac{\pi}{2} \right]$, by referring to Jensen's inequality, we have
\begin{equation}
\begin{aligned}
\sum_{n=1}^{\frac{N_{zc}-1}{2}} \frac{1}{\tan(\frac{\pi n}{N_{zc}})} & \leq \int_{\frac{1}{2}}^{\frac{N_{zc}}{2}} \frac{1}{\tan(\frac{\pi n}{N_{zc}})} \mathrm{d}n = Q,
\end{aligned}
\label{eq:upper_bound3}
\end{equation}
where
\begin{equation}
\begin{aligned}
Q=-\frac{N_{zc}}{\pi} \ln \left[ \sin \left( \frac{\pi}{2N_{zc}}   \right)  \right].
\end{aligned}
\label{eq:Q}
\end{equation}
Up to now, it suffices to derive the upper bound of inter-preamble interference in Case 2 $M_2(\beta^{u^\prime},k)$ based on (\ref{eq:upper_bound1}) and (\ref{eq:upper_bound2})-(\ref{eq:Q}) as follows.
\begin{equation}
\begin{aligned}
M_2(\beta^{u^\prime},k)= \min\left\{(\frac {k\beta^{u^\prime}}{N_{zc}})^2,\frac{k^2(\beta^{u^\prime}+2Q )^2}{{N_{zc}}^3} \right\}.
\end{aligned}
\label{eq:pp_bound}
\end{equation}

\subsubsection{The upper bound of interference in Case 3}\label{sec:case3_analysis}
\
\newline
\indent
Based on (\ref{eq:cal frac_pdp}) and (\ref{eq:32}), PDP computation result between the sum of the two interfering sequences and a ZC sequence $\{x(n,r_b^u)\}$ in Case 3 can be expressed as
\begin{equation}
	\begin{aligned}
		&PDP(m,r_b^u)\\
		& \leq \left [ \frac{1}{\sqrt{N_{zc}}^3}\sum_{n=0}^{N_{zc}-1} \left({k_1} \vert \widehat \rho_{\alpha_1^{u^\prime},\beta_1^{u^\prime}}(n) \vert \right. \left. + {k_2} \vert \widehat \rho_{\alpha_2^{u^\prime},\beta_2^{u^\prime}}(n) \vert \right) \right]^2 \\
        &<\frac{1}{N_{zc}^3}  \left [k_1\beta_1^{u^\prime}+2k_1 \sum_{n=1}^{\frac{N_{zc}-1}{2}} \frac{1}{\tan(\frac{\pi n}{N_{zc}})}+k_2 \beta_2^{u^\prime}  \right.\\
        &\left.   + 2k_2 \sum_{n=1}^{\frac{N_{zc}-1}{2}} \frac{1}{\tan(\frac{\pi n}{N_{zc}})} \right]^2 \\
        &\leq\frac{1}{N_{zc}^3}[ k_1(\beta_1^{u^\prime}+2Q)+k_2(\beta_2^{u^\prime} +2Q) ]^2 \\
		& = [\sqrt{M_2(\beta_1^{u^\prime},k_1)}+\sqrt{M_2(\beta_2^{u^\prime},k_2)}]^2,
	\end{aligned}
	\label{eq:interference_2}
\end{equation}
where $k_1$ and $k_2$ are the amplitude ratio of the interfering sequence 1 and sequence 2 to a ZC sequence $\{x(n,r_b^u)\}$ generated by satellites, respectively.
Two windowing sequence  $\rho_{\alpha_1^{u^\prime},\beta_1^{u^\prime}}$ and $\rho_{\alpha_2^{u^\prime},\beta_2^{u^\prime}}$ satisfy $\beta_1^{u^\prime} + \beta_2^{u^\prime} = N_{zc}$ as shown in Fig.~\ref{fig5}.
Then, the upper bound of interference in Case 3 $M_3(\beta^{u^\prime},k_1,k_2)$ can be given by
\begin{equation}
	M_3(\beta^{u^\prime},k_1,k_2) = [\sqrt{M_2(\beta^{u^\prime},k_1)}+\sqrt{M_2(N_{zc}-\beta^{u^\prime},k_2)}]^2.
   \label{eq:upper_bound3}
\end{equation}
Moreover, Case 2 is a specific situation of Case 3 with one of interfering sequence being zero sequence, where $k_1=0$ or $k_2=0$.

\subsubsection{Interference analysis for general multi-UE case}\label{sec:multi-UE case}
\
\newline
\indent
With the above analysis, we further extend them to general multi-UE access.
Assuming that the interference term in (\ref{eq:interference_term}) contains multiple incomplete ZC sequences from $U'$ UEs, the PDP computation between the sum of these ZC sequences and a ZC sequence $\{x(n,r_b^u)\}$ can be written as follows.
\begin{equation}
	\begin{aligned}
		PDP(m,r_b^u)
		& \leq \left [\sum_{u^\prime=1}^{U^\prime} \frac{1}{\sqrt{N_{zc}}^3}\sum_{n=0}^{N_{zc}-1} \left({k_1^{u^{\prime}}} \vert \widehat \rho_{\alpha_1^{u^\prime},\beta_1^{u^\prime}}(n) \vert \right. \right. \\
		&\left. \left. + {k_2^{u^{\prime}}}\vert \widehat \rho_{\alpha_2^{u^\prime},\beta_2^{u^\prime}}(n) \vert \right) \right]^2 \\
& \leq [\sum_{u^\prime=1}^{U}\sqrt{M_2(\beta_1^{u^\prime},k_1^{u^{\prime}})}+\sqrt{M_2(\beta_2^{u^\prime},k_2^{u^{\prime}})}]^2 \\
& = [\sum_{u^\prime=1}^{U} \sqrt{M_3(\beta^{u^\prime},k_1^{u^{\prime}},k_2^{u^{\prime}})} ]^2,
	\end{aligned}
	\label{eq:upper_bound4}
\end{equation}
where $k_1^{u^{\prime}}$ and $k_2^{u^{\prime}}$ are the amplitude ratio of the interfering sequence 1 and 2 of UE $U^\prime$ to the ZC sequence  $\{x(n,r_b^u)\}$, respectively.

\subsubsection{The probability that partial-period cross-correlation occurs }\label{sec:probability_analysis}
\
\newline
\indent
At last, we first study the probability that partial-period cross-correlation operations occur in module 2 shown in Fig.~\ref{fig5} based on the preamble format with fixed cascading order of ZC sequences in time-domain.
As shown in Fig.~\ref{fig5}, we can observe that a  partial-period cross-correlation interference must happen if the TA difference between UE $u$ and $u^\prime$ satisfies that$\mod(T^u-T^{u^\prime},T_{zc}) \neq 0$, and $T_{zc}$ represents the duration of a ZC sequence in time domain.
We ignore the situation that$\mod(T^u-T^{u^\prime}, T_{zc})= 0$ in the following since the probability of it occurring is extremely low.
Define $Pr_u(j)$ as the probability of the $j$-th subsequence $\{y_{u,j}^\prime(n)\}$ of UE $u$ in module 2 containing incomplete ZC sequences from another UE $u^\prime$, which means that partial-period cross-correlation interference occurs.

\begin{figure}[!t]
	\center
	\includegraphics[width=2.5in]{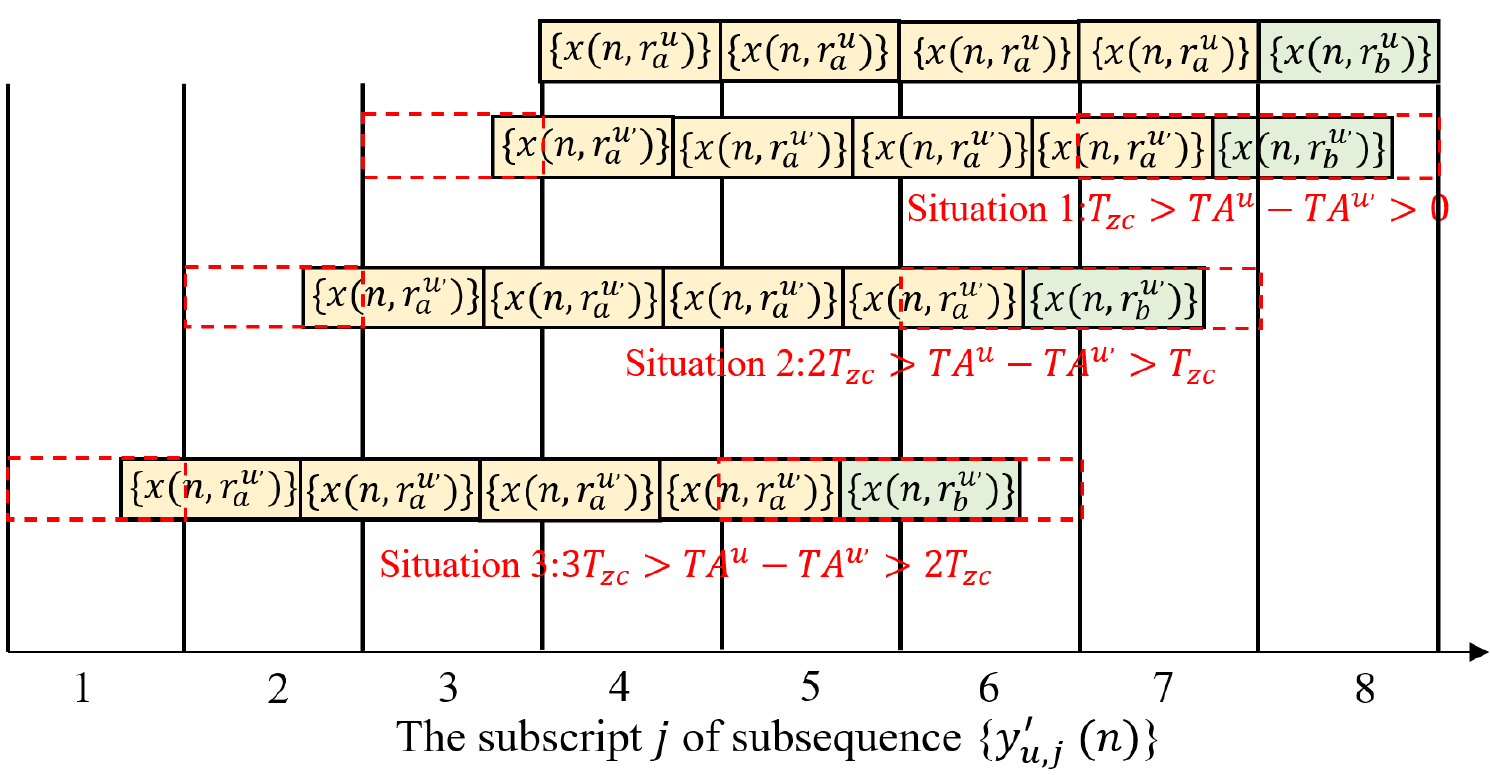}
	\caption{An example about all potential interference cases of UE $u$ when $0<{TA}^u-{TA}^{u^\prime} <3T_{zc}$, $Z_l=5$, $G_l=3$, $b^u =5$ and $TA^u=K_f^u+3T_{zc}$.}\label{fig6}
\end{figure}

Assume that TA difference is uniformly distributed within $[0,G_lT_{zc}]$, where $G_l=\lceil \frac{2(T_{u,s}^{max}-T_{u,s}^{min})}{T_{zc}} \rceil$ and $T_{u,s}^{max}$, $T_{u,s}^{min}$ represent the maximal and minimal propagation delay between satellite $s$ and UEs in a beam cell, respectively.
Define $b^u$ as the position of ZC sequence with root $r_b^u$ in preamble and $b^u \in \{1,2,...,Z_l\}$, where $Z_l$ is the number of ZC sequences in a preamble.
In the following, we take the Fig.~\ref{fig6} as an example to illustrate all potential interference cases and probability between two UEs, where we set $0<{TA}^u-{TA}^{u^\prime} <3T_{zc}$ and $K_i^u=3$.
It can be observed that UE $u^\prime$ makes three subsequences $\{y_{u,j}^\prime(n)\}$ of UE $u$ suffer partial-period cross-correlation interference in all situations.
Moreover, only $\{y_{u,1}^\prime(n)\}$,...,$\{y_{u,G_l}^\prime(n)\}$ and $\{y_{u,b_u}^\prime(n)\}$,...,$\{y_{u,b_u+G_l}^\prime(n)\}$ may suffer partial-period cross-correlation interference, since ZC sequence $\{x(n,r_b^{u^\prime})\}$ and the first ZC sequence $\{x(n,r_a^{u^\prime})\}$ from the preamble of UE $u^\prime$ are distributed from subsequences $\{y_{u,b_u}^\prime(n)\}$ to $\{y_{u,b_u+G_l}^\prime(n)\}$ and from $\{y_{u,1}^\prime(n)\}$ to $\{y_{u,G_l}^\prime(n)\}$, respectively.
For example, only $\{y_{u,1}^\prime(n)\}$,...,$\{y_{u,3}^\prime(n)\}$ and  $\{y_{u,5}^\prime(n)\}$, ...,$\{y_{u,8}^\prime(n)\}$ may perform partial-period cross-correlation operations in Fig.~\ref{fig6}.

If $G_l<j < b^u$ , subsequence $\{y_{u,j}^\prime(n)\}$ in all Situations consists of complete ZC sequences and its cyclic shifted sequences that only produce full-period cross-correlation operations.
When $j=b^u$ ar $j = b^u+G_l$, only preambles from other UEs with arrival time difference distributed within $[(G_l-1)T_{zc},G_lT_{zc}]$ or $[0,T_{zc}]$ lead to incomplete ZC sequences in $\{y_{u,j}^\prime(n)\}$.
When $1 \leq j \leq G_l$, only preambles from other UEs with arrival time difference distributed within $[(j-G_l)T_{zc},(j-G_l+1)T_{zc}]$ lead to incomplete ZC sequences in $\{y_{u,j}^\prime(n)\}$.
For example, we can observe that subsequence $\{y_{u,3}^\prime(n)\}$ and $\{y_{u,8}^\prime(n)\}$ only perform partial-period cross-correlation operations in Situation 1.
In addition,
when $b^u+1 \leq j \leq b^u+G_l-1$, each subsequence $\{y_{u,j}^\prime(n)\}$ performs partial-period cross-correlation operation when TA difference is distributed within
$[(b^u+G_l-j-1)T_{zc} , (b^u+G_l-j+1)T_{zc}]$.
For example, subsequence $\{y_{u,6}^\prime(n)\}$ suffers partial-period cross-correlation interference in Situation 2 and 3, corresponding to TA difference distributed within $[0,2T_{zc}]$.

Under the assumption that the arrival time difference of UE preambles follows uniform distribution, the probability that partial-period cross-correlation occurs in the PDP computation of a subsequence $\{y_{u,j}^\prime(n)\}$ is proportional to the length of the TA difference.
Then, we have
\begin{equation}\label{G_l}
Pr_u(j)=\left\{\begin{array}{ll}
\frac{1}{G_l}, & 1\leq j \leq G_l, \;j=b^u \,or \,b^u+G_l , \\
\frac{2}{G_l}, &b^u +1 \leq j\leq b^u+G_l-1, \\
0, &otherwise.
\end{array}\right.
\end{equation}
Note that, the sum of $Pr_u(j)$ is equal to 3, which means that each interference UE $u^\prime$ causes three subsequences $\{y_{u,j}^\prime(n)\}$ to perform partial-period cross-correlation operations.

%---------------------------------
\subsection{Flexible preamble format design}\label{sec:proposed_preamble_design2}
%---------------------------------

As mentioned in Section \ref{sec:TA_estimation}, TA of each UE is estimated based on the autocorrelation property of ZC sequence and peak detection.
However, due to partial-period cross-correlation operations, the amplitude of $PDP(m,r_b^u)$ of the interfering term may be larger than the autocorrelation peak, which generates several pseudo-peaks and hence misleads the estimation of $K_i^u$.
In (\ref{eq:upper_bound4}), the upper bound of inter-preamble interference in each subsequence $\{y_{u,j}^\prime(n)\}$ is shown to be proportional to the square of the number of interfering UEs who bring partial-period cross-correlation operations in it.
Therefore, the performance of current preamble design rapidly drops in multi-UE case.
Hence, it is crucial to reduce adverse impacts of performing partial-period cross-correlation operations in PDP computation for each subsequence $\{y_{u,j}^\prime(n)\}$.
To tackle the above issue, we propose a flexible preamble format design based on ZC sequences with two roots advised by 3GPP in this subsection.
Our proposal further includes preamble design and preamble format duration design. Moreover, it will be shown that differential power allocation among ZC sequences with different roots significantly affect inter-preamble interference.

\subsubsection{Preamble design }\label{sec:preamble_design}
\begin{figure}[!t]
	\center
	\includegraphics[width=2.6in]{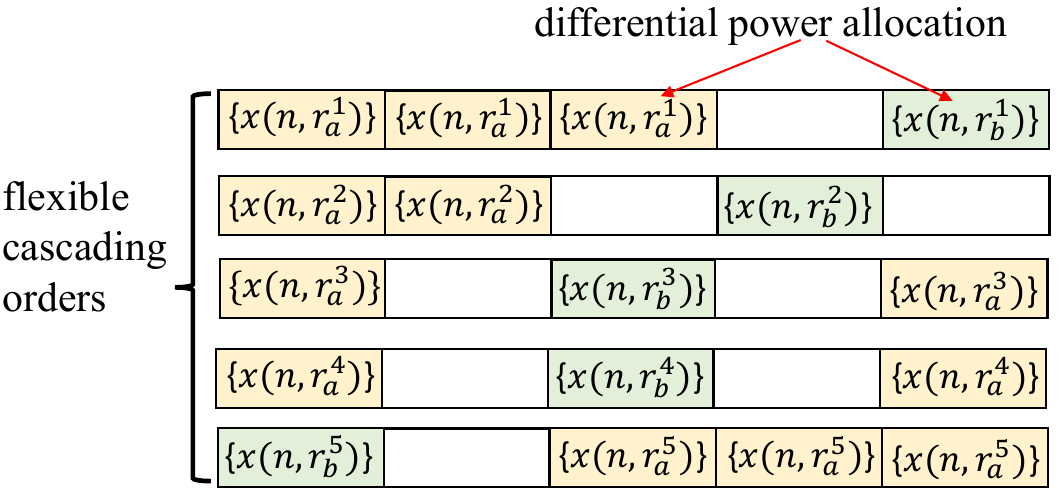}
	\caption{The proposed preamble design based on ZC sequences with two roots.}\label{fig7}
\end{figure}

\
\newline
\indent
Flexible preamble design is proposed for 5G LEO networks as shown in Fig.~\ref{fig7}, aiming to improve the accuracy of preamble detection and TA estimation.
The core idea of our proposal can be concluded as following.
\begin{itemize}
	\item \emph{Flexibly cascading ZC sequences with different roots:} reduce the probability that subsequence  $\{y_{u,j}^\prime(n)\}$ consists of massive incomplete ZC sequences.
	\item \emph{Allocating differential power levels among ZC sequences:} reduce the impact of once partial-period cross-correlation operation.
\end{itemize}

Particularly, flexible cascading order of ZC sequences can fully utilize the time domain resource of preamble and does not change its duration length.
Moreover, the differential power allocation strategy among ZC sequences can improve the amplitude of the detected autocorrelation peaks.
When the amplitude of a ZC sequence is increased by $k$-fold and others keep unchanged, the autocorrelation peak increases by $k^2$-fold.
Meanwhile, the interference from another preamble does not increase by the same level, and this is why differential power allocation among ZC sequences can contribute to less pseudo-peaks.
However, with such design, UEs are required to perform rapid power adjustment in time domain, which is challenging for hardware capability.
Therefore, as shown in Fig.~\ref{fig7}, both sides of the high power ZC sequence are further replaced by two sequences with amplitude 0, which is equivalent to no data transmission.

Note that our proposal can also be used to enhance the other two optional preamble formats advised by 3GPP in Section~\ref{sec:preamble_option}.
For example, we can change the amplitude of the first ZC sequence in preamble for Option-1.
In this case, the satellite estimates $K_i^u$ by detecting the first ZC sequence.
As for Option-2, we can also change the amplitude of a ZC sequence scrambled by an M sequence and then change its position flexibly in preamble.
At this time, the satellite estimates $K_i^u$ by detecting the ZC sequence with higher power and an additional descrambling step is also required.

\subsubsection{Preamble format duration design }\label{sec:duration_design}
\
\newline
\indent
As mentioned before, 5G preamble format proposed in 3GPP standard TS 38.221 of Rel. 17~\cite{38.221} consists of CP, preamble and guard time.
Note that CP is utilized to alleviate inter-symbol interference when the sum of maximal arrival time difference of preambles and channel delay spread is less than the duration of CP.
In terrestrial 5G systems, the base station first discards samples corresponding to CP in the preamble detection procedure and then performs correlation operations.
Due to the significant arrival time difference of preambles in LEO networks, CP duration reaches a millisecond level under the above set, causing significant overhead.
In addition, the transmitted signals of CP are a replica of the last symbols of the preamble.
Therefore, CP length in preamble format can be set to zero\cite{huawei,zte}, and more time resource are allocated for preamble transmitting, which means increasing the number of ZC sequences in preambles.
Meanwhile, the preamble duration $T_P$ must be larger than the maximal difference of TA.
In TDD mode, random access procedure usually begins at uplink pilot time slot, which means that only  TA estimation error $T_e^u>0$ may bring interference to normal PUSCH transmission of other UEs.
To avoid this interference, guard time must be larger than the maximal $T_e^u$ among UEs.
Define $T_{CP}$, $T_{GT}$, and $T_{P}$ as the duration of CP, guard time and preamble, respectively, which satisfy
\begin{equation}
\left\{
             \begin{array}{lr}
             T_{CP}=0, & \\
             T_{P} > T_{GT} \geq \max (T_e^u).
             \end{array}
\right.
\end{equation}

\subsection{Discussion on the performance gain of our proposal}

In this subsection, we discuss the advantages of our preamble format design brought by differential power allocation among ZC sequences and flexible cascading order of them.

\subsubsection{The benefit of differential power allocation}\label{sec:benefit1}
\
\newline
\indent
If the amplitude of the ZC sequence with root $r_b^u$ is increased by $k$-fold relative to that of ZC sequence with root $r_a^u$, the autocorrelation peak in module 2 is $k^2$ according to (\ref{eq:cal frac_pdp}) and (\ref {eq:cor_cal}).
Meanwhile, the amplitude ratio of an interfering sequence to a ZC sequence in Case 3 becomes $k_1k$ , and another one keeps unchanged.
Based on (\ref{eq:pp_bound}) and (\ref{eq:upper_bound3}), the ratio of interference upper bound $M_3(\beta,k_1,k_2)$ to autocorrelation peak $k^2$ in our proposal can be  expressed as
\begin{equation}
	\begin{aligned}
		&R(k,\beta,k_1,k_2)= \frac{[k\sqrt{M_2(\beta,k_1)}+\sqrt{M_2(N_{zc}-\beta,k_2)}]^2}{k^2}    \\ &=[\sqrt{M_2(\beta,k_1)}+\frac{\sqrt{M_2(N_{zc}-\beta,k_2)}}{k}]^2.
	\end{aligned}
	\label{eq:normalized_I}
\end{equation}
Note that a smaller $R(k,\beta,k_1,k_2)$ can contribute to a more accurate TA estimation.
From the above equation, it is observed that differential power allocation among ZC sequences with different roots, i.e., $k>1$,
can decrease $R(k,\beta,k_1,k_2)$.
However, as $k$ approaches infinite, $R(k,\beta,k_1,k_2)$ approaches $M_2(\beta,k_1)$, which means that the benefit brought by increasing $k$ is limited.

\subsubsection{The benefit of flexible cascading}\label{sec:benefit2}
\
\newline
\indent
In this part, we compare the proposed preamble design with a fixed preamble design, where the position of ZC sequence with root $r_b^u$ is fixed.
The probability of performing partial-period cross-correlation under fixed preamble design is discussed in Section \ref{sec:interference_analysis}, which is denoted as $Pr_u(j)$.
In the following, we further explore the probability that partial-period cross-correlation occurs in the PDP computation of subsequence $\{y_{u,j}^\prime(n)\}$ under our preamble design, and the probability is denoted as $Pr_u^\prime(j)$ with $j$ being the subscript of $\{y_{u,j}^\prime(n)\}$.

Recall that $b^u$ is the position of the ZC sequence with root $r_b^u$ in preamble and is an integer in the range $\{1,2,...,Z_l\}$, where $Z_l$ is the number of ZC sequences in preamble.
Assuming the arrival time of UE $u$'s preamble is uniformly distributed within $[0,G_l T_{zc}]$.
At the beginning, we fix $b^u$ and then $Pr_u^\prime(j)$ is equivalent to $Pr(j)$ in this case.
Owning to flexible cascading order among ZC sequences, $b^u$ is uniformly distributed in $\{1,2,...,Z_l\}$.
Therefore, $Pr_u^\prime(j)$ can be derived based on the joint distribution function of two uniform distributions, which is expressed as
\begin{equation}\label{Z_l}
Pr_u^\prime(j)=\left\{\begin{array}{ll}
\frac{1}{G_l}, & j=1,Z_l+G_l , \\
\frac{1}{G_l}+\frac{2(j-1)}{G_l Z_l}, &2\leq j \leq G_l, \\
\frac{2}{Z_l}, &G_l <j\leq Z_l\\
\frac{1}{G_l}+\frac{2(G_l+Z_l-j)}{G_l Z_l},&otherwise.
\end{array}\right.
\end{equation}
In (\ref{Z_l}), when $j\leq G_l$, $Pr_u^\prime(j)$ increases monotonically with $j$, and $Pr_u^\prime(j)$ decreases monotonically with $j$ when $Z_l \leq j$. Therefore, $\max (Pr_u^\prime(j))=\frac{2}{Z_l}$.
In addition, $\max(Pr_u(j))=\frac{2}{G_l}$ in (\ref{G_l}) and $Z_l > G_l$ due to $Z_lT_{zc} = T_p >T_{GT} \geq G_lT_{zc}$.
Then, it can be concluded that $\max(Pr_u(j)) > \max(Pr_u^\prime(j))$, which means that our proposal reduces the probability that subsequence  $\{y_{u,j}^\prime(n)\}$ consists of massive incomplete ZC sequences than the preamble design with fixed cascading orders.
Moreover,  the expression of $\max(Pr_u^\prime(j))$ points out that a longer duration length of the proposed preamble can also improve TA estimation performance.

%---------------------------------
\subsection{Preamble detection complexity}
%---------------------------------
Note that our proposed preamble format design does not change the estimation process of $K_f^u$, and hence our proposal has the same detection complexity when estimating $K_f^u$ compared with other preamble format design options in Section~\ref{sec:preamble_option}.
For the fixed preamble format design, the maximal arrival time difference of preambles is $G_lT_{zc}$, which means that satellites only require to perform $G_l$ PDP computations to obtain $K_i^u$.
In addition, our proposal requires calculating the PDP for all subsequences $\{y_{u,j}^\prime(n)\}$ given in (\ref{eq:y_si}), which means that $Z_l+G_l$ PDP computations are needed.
Based on these analysis, the preamble detection complexity at the satellite with our proposal is increased by $\frac{Z_l}{G_l}$ compared with fixed format design.

%---------------------------------
\subsection{The application of our preamble format design in practice}
%---------------------------------
In this subsection, we provide two possible ways of applying our proposed preamble format design.
In the first manner, the satellite indicates the preamble length $Z_l$, the roots of each preamble, the amplitude ratio $k$ of the ZC sequence with $r_b^u$ to the ZC sequence with $r_a^u$ and the position of the ZC sequence with root $r_b^u$ in preamble $b^u$ via common physical downlink shared channel, whose detection is guided by SSB information.
As for the calculation of actual TA, it is given by
\begin{equation}
	T^u=T_{pre}^u +(j^*-b^u) T_{zc} + K_f^u,
	\label{eq:acutal_TA cal2}
\end{equation}
where $T_{pre}^u$ is the time pre-compensation value, $K_f^u$ is the fractional part of TA estimation, and $j^*$ is the subscript of subsequence $\{y_{u,j}^\prime(n)\}$ acquired in the estimation procedure of $K_i^u$.
In the second manner, UEs select $b^u$ by themselves and UEs should transmit $b^u$ via PUSCH in both 2-step and 4-step random access procedures for the satellite to calculate $T^u$.
In addition, our proposal can be applied to both satellites with regenerative payloads and with transparent payload scenarios, where the gateway station on the ground performs time-frequency compensation for the feeder link.
For the satellite with transparent payload, TA equals the sum of the result in (\ref{eq:acutal_TA cal2}) and time compensation value of the feeder link.

Our proposal can be applied to several frequency bands.
In L, S and X band, due to low operation frequency and precise equipment components,
frequency offset caused by local oscillators may be less than half of SCS, which means that UEs can take the value of the measured downlink frequency offset as the values of uplink frequency pre-compensation.
Then, the proposed preamble design can also be applied directly.
In high-frequency bands, UEs can perform a coarse time-frequency pre-compensation based on the Algorithm 1 in Section \ref{sec:time-frequency pre-compensation}, and then transmit the proposed preamble to satellites.

%---------------------------------
\section{Simulation Results and Analysis}\label{sec:sim}
%---------------------------------

In this section, extensive simulations are presented to evaluate the performance of the proposed TA estimation approach.
We first investigate the impacts of operation frequency on time-frequency pre-compensation, and then verify the superiority of the proposed preamble format design by demonstrating the missed detection rate of preamble and TA estimation error.

%---------------------------------
\subsection{Parameter setting and performance metrics}
%---------------------------------
\begin{table}
\centering
\caption{Simulation Parameter Setting}
\label{tab:1}
\scriptsize
\begin{tabular}{l l} \toprule
\textbf{Parameters} & \textbf{Value} \\
\midrule
The number of LEO satellite orbit & 20 \\
The number of satellites in an orbit & 15\\
Orbit altitude & 1000 km\\
Orbit inclination &  $53^{\circ}$ \\
Frequency offset caused by local oscillators & $[0,5\times10^{-7}]f_s$ \\
Downlink frequency offset measurement error  & [-1.2,1.2] kHz \\
The duration of a time slot  & 0.5 ms \\
SCS of PRACH & 30 kHz \\
$N_{zc}$ & 571 \\
The range of roots &[1,570] \\
$T_{P}$  &  0.733 ms \\
$T_{GT}$   &  0.267 ms \\
$f_s$  & 27 GHz         \\
Channel model       &  AWGN  \\
\bottomrule
\end{tabular}
\end{table}

\begin{figure*}[htpb]
    \centering
        \subfigure[]
		{
                \includegraphics[width=2.2in]{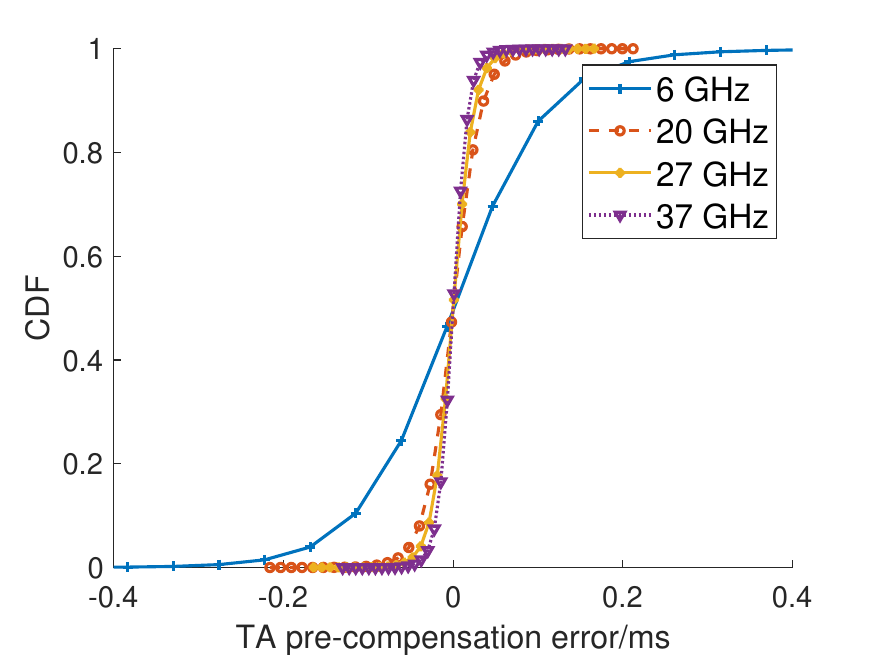}
        }
        \subfigure[]
		{
                \includegraphics[width=2.2in]{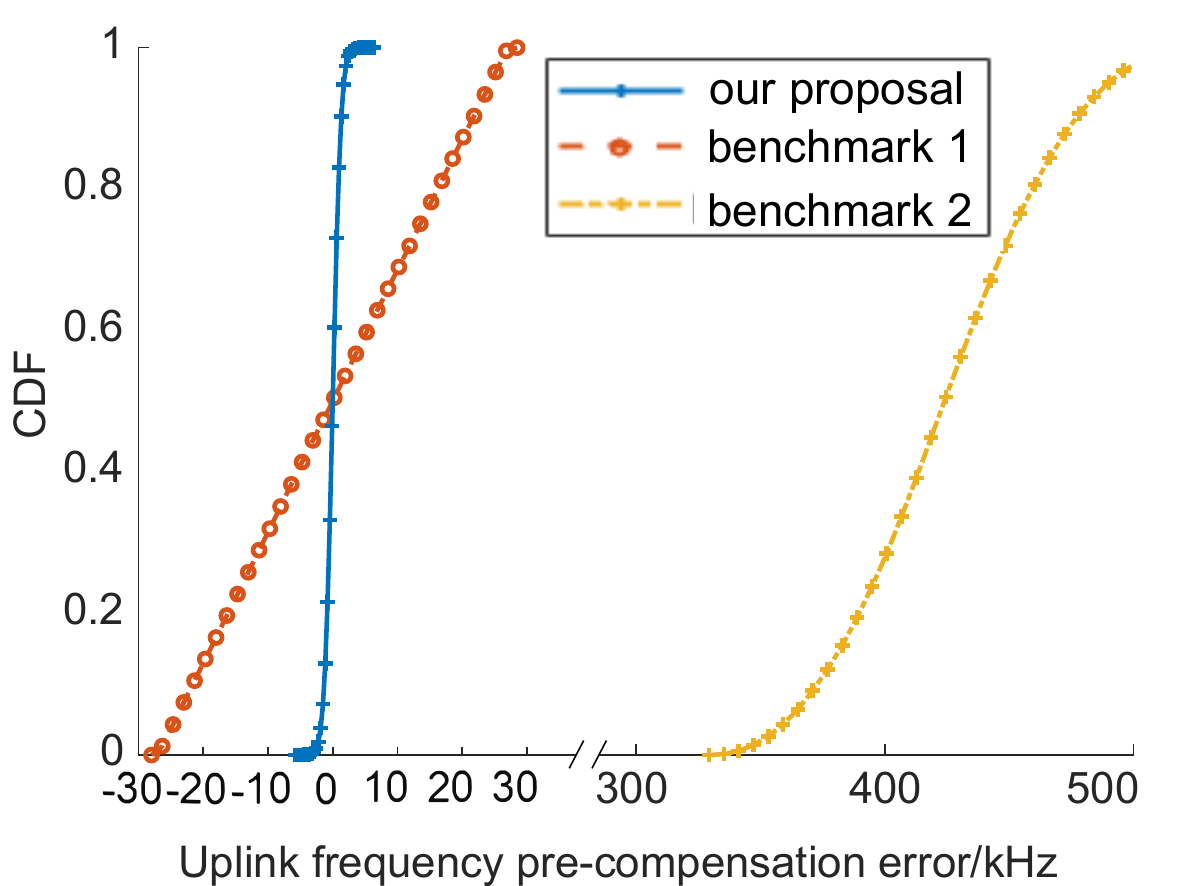}
        }
        \subfigure[]
		{
                \includegraphics[width=2.2in]{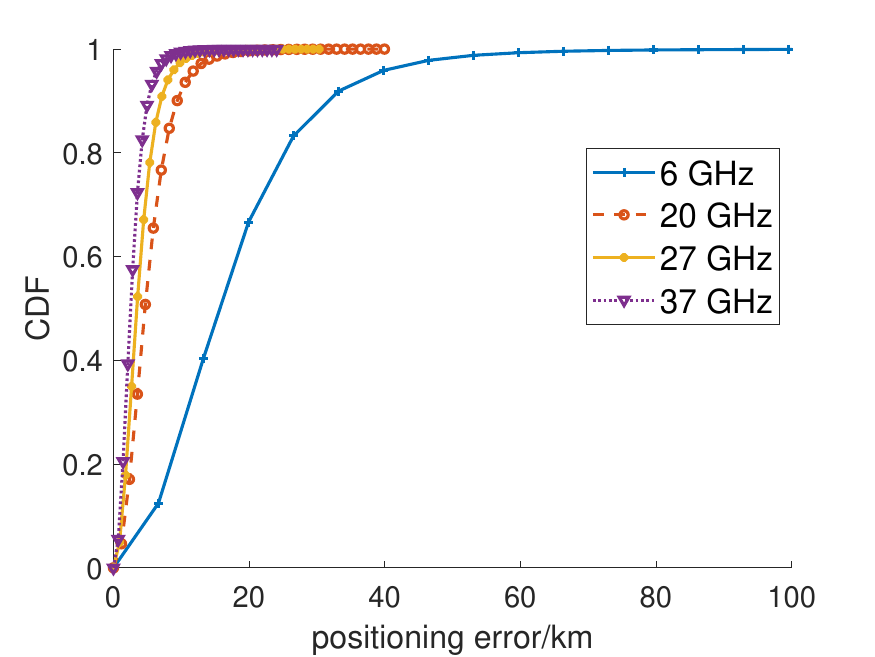}
        }
    \caption{The figures are the result of the performance of Algorithm~\ref{alg:1}, where (a) presents CDF of TA pre-compensation error, (b) is the CDF of uplink frequency pre-compensation error, and (c) shows the CDF of positioning error.}\label{fig8}
\end{figure*}

In simulations, an LEO constellation with 300 satellites evenly distributed across 20 orbits is generated by System Tool Kit (STK) simulator, which ensures more than 3 satellites is always in the view of any UE on the Earth.
The inclination and altitude of each satellite orbit are set to $53^{\circ}$ and $1000$ km, respectively.
Maximum frequency measurement error in downlink synchronization procedure is 1.2 kHz and the frequency offset caused by local oscillators follows uniform distribution within $[0,5\times10^{-7}]$ of carrier frequency.
In addition, we set the subcarrier spacing of PRACH to 30 kHz and
UE preamble is generated by cascading multiple ZC sequences in time domain with $N_{zc}=571$.
The operation frequency $f_s$ of satellites is 27 GHz.

Since TA estimation accuracy highly depends on the preamble detection procedure, we mainly evaluate the preamble detection performance.
The performance metrics of preamble detection include false alarm rate and missed detection rate\cite{metric}.
The false alarm rate is defined as the rate that the satellite detects a dummy preamble, which is not transmitted by UEs.
By referring to \cite{metric,missed_detection_rate2}, the missed detection rate is defined as the rate that the satellite does not correctly detect a transmitted preamble, including two error cases as follows.
\begin{itemize}
	\item  The transmitted preamble is not detected.
    \item  The preamble is detected but the corresponding TA estimation error exceeds the duration length of CP of PUSCH.
\end{itemize}
Moreover, we set the threshold of peak detection based on 1$\%$ false alarm rate.
The uplink channel is an additive white gaussian noise (AWGN) channel~\cite{38.811}.
Other main parameters are summarized in Table~\ref{tab:1}.

%---------------------------------
\subsection{Performance of the proposed time-frequency pre-compensation method}
%---------------------------------

In this part, we focus on evaluating the performance of our proposed time-frequency pre-compensation method in Algorithm~\ref{alg:1}.
Meanwhile, frequency pre-compensation only (benchmark 1)\cite{COE} and no pre-compensation (benchmark 2)
are compared with our proposal, and benchmark 1 directly uses downlink frequency offset measurement to compensate
uplink frequency offset.
Considering that the timing offset and frequency offset of UEs with high elevation angles are relatively small, the elevation angles of UEs are randomly generated from $[20^{\circ}, 70^{\circ}]$.
In addition, each UE measures the downlink frequency offset by detecting SSBs from three satellites for only one time.

Fig.~\ref{fig8} (a) shows the TA estimation performance of the proposed method in the initial cell search phase, where the maximal TA difference among UEs is set to 0.75 ms.
It can be observed that the performance of Algorithm~\ref{alg:1} is sensitive to operation frequency and operation frequency of 27 GHz can guarantee TA pre-compensation error does not exceed the range of $[-0.1, 0.1]$ ms.
In 5G random access protocol, TA command (TAC) with 12 bits is adopted to instruct UEs to adjust uplink timing by maximal value of 0.5 ms\cite{common_TA3}, which well supports pre-compensation error of $[-0.1, 0.1]$ ms.
This means that our proposed time-frequency pre-compensation method can be applied without modifying TAC.
Meanwhile, when the beam cell diameter is 170 km and the elevation angle at the center of the beam is $45^{\circ}$, our proposal can release 0.55 ms time domain resource for PUSCH transmission.

Next, we evaluate the effect of Algorithm~\ref{alg:1} on uplink frequency pre-compensation error when satellite operation frequency is 27 GHz.
According to Fig.~\ref{fig8} (b), the probability that residual frequency offset is less than 3.7 kHz reaches $99.865\%$.
However, if we take the measured downlink frequency offset as the uplink frequency pre-compensation value, the residual frequency offset  may reach 30 kHz.
Moveover, if no frequency pre-compensation method is implemented, uplink CFO can reach 500 kHz.
The positioning performance of our proposal is illustrated in Fig.~\ref{fig8} (c).
With the operation frequency increasing, our proposal achieves more accurate positioning results, which contributes to decreasing the time-frequency pre-compensation error.
When the satellite operation frequency is 27 GHz, positioning error is distributed within [0.0069, 30.3308] km.

\begin{figure}[htpb]
	\center
	\includegraphics[width=2.5in]{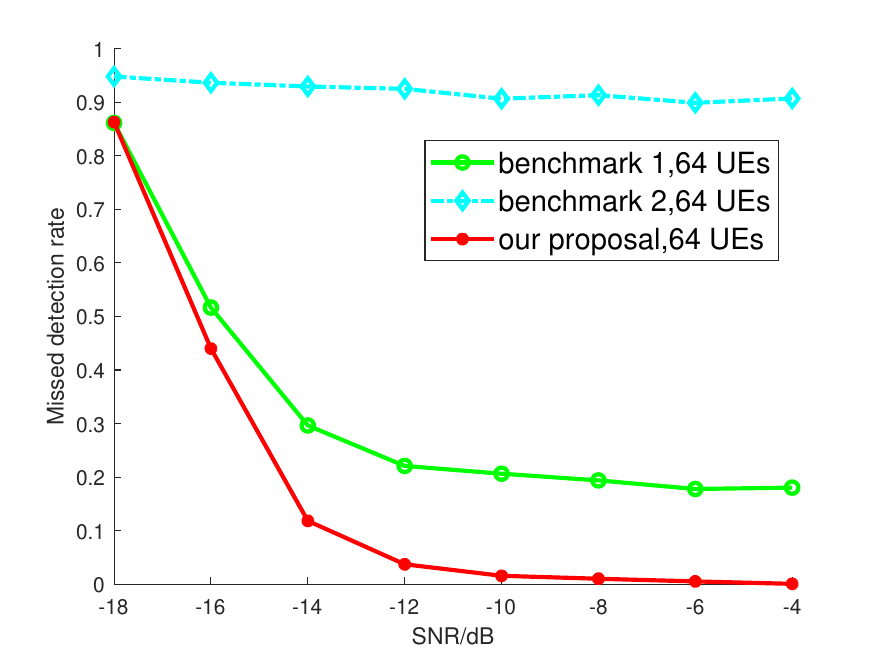}
	\caption{The missed detection rate of the proposed preamble design under different frequency pre-compensation methods.}\label{fig9}
\end{figure}

Fig.~\ref{fig9} illustrates the preamble missed detection rates under different frequency pre-compensation methods.
In simulation, we set the maximal arrival time difference and uplink CFO is consistent with that in Fig.~\ref{fig8}.
For simplicity and fair comparisons, we neglect serial interference cancellation and the influence of preamble collision among UEs\cite{zhenli,zte}.
In addition, the amplitude ratio $k$ in our preamble design is set to 2.
When UEs number is 64, the proposed pre-compensation method can achieve $1\%$ missed detection rate and at least $17.95\%$ performance improvement compared with other benchmarks
that do not consider the impact of frequency offset incurred by local oscillators.

%---------------------------------
\subsection{Performance of the proposed preamble format design}
%---------------------------------

In this subsection, we first show the negative effect of partial-period cross-correlation operations on TA estimation.
Then, the missed detection rate and TA estimation error with our preamble format design is evaluated.
In simulation, the maximal arrival time difference and uplink frequency offset of preambles are set to 0.2 ms and 3 kHz, respectively.
Later on, we compare the performance of our proposal with other benchmarks that are based on preamble format Option-3 in Section~\ref{sec:preamble_option}, in terms of TA parameter $K_f^u$ estimation. The benchmark schemes include:
\begin{itemize}
	\item \emph{Preamble design A:} original preamble format Option-3\cite{2root}.
	\item \emph{Preamble design B:}  preamble format Option-3 with differential power allocation strategy.
	\item \emph{Preamble design C:}  preamble format Option-3 with flexible cascading orders of ZC sequences.
\end{itemize}

To verify the generalization of our preamble format design idea, we further evaluate the missed detection rates under preamble formats Option-1 and Option-2 in Section~\ref{sec:preamble_option} as follows with 64-UE access.
\begin{itemize}
	\item \emph{Preamble design D:} original preamble format Option-1\cite{38.821}.
    \item \emph{Preamble design E:} preamble format Option-1 enhanced by our proposal.
	\item \emph{Preamble design F:} original preamble format Option-2 based on M sequence scrambling\cite{weighted}.
    \item \emph{Preamble design G:} preamble format Option-2 enhanced by our proposal.
    \item \emph{Preamble design H:} preamble format Option-3 enhanced by our proposal.
\end{itemize}
Other simulation parameters are consistent with the previous subsection.

\begin{figure}[!t]
\center
\includegraphics[width=2.5in]{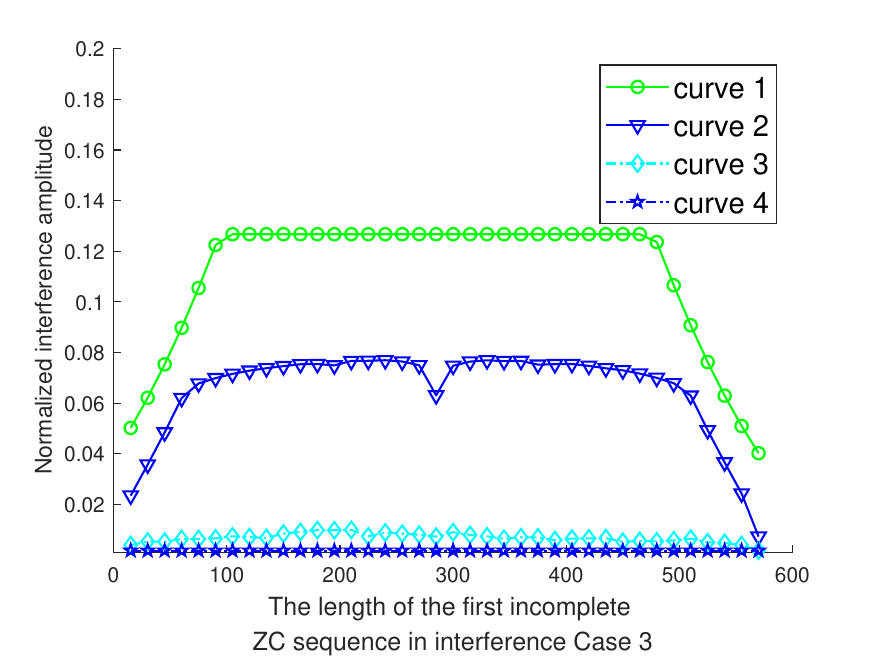}
\caption{Comparison among the derived upper bound final expression of interference in Case 3 (curve 1), the derived upper bound expression of interference based on the second term equation (\ref{eq:upper_bound2})  in Case 3 (curve 2), the real interference in Case 3 (curve 3) and interference caused by full-period cross-correlation operation (curve 4).}\label{fig10}
\end{figure}

Fig.~\ref{fig10} provides a comparison among the derived upper bound of interference with $k=1$ in Case 3 (curve 1,2) discussed in Section~\ref{sec:interference_analysis}, the real interference in Case 3 (curve 3) and interference caused by full-period cross-correlation operations (curve 4), where curve 1 is obtained based on the third term of (\ref{eq:upper_bound2}) and curve 2 is obtained based on the second term of (\ref{eq:upper_bound2}), respectively.
Meanwhile, amplitudes of  all curves are normalized with respect to the autocorrelation peak of ZC sequences.
The figure shows that the derived upper bound of inter-preamble interference in (\ref{eq:upper_bound3}) is reasonable, and the interference caused by partial-period cross-correlation operations in PDP computation is dominant in multi-UE case.
The upper bound of expression can be used to select the number of ZC sequences in preamble format that ensures the probability of successful access, under the given length of a ZC sequence, missed detection rate and the access number of UEs.

\begin{figure}[!t]
	\center
	\includegraphics[width=2.5in]{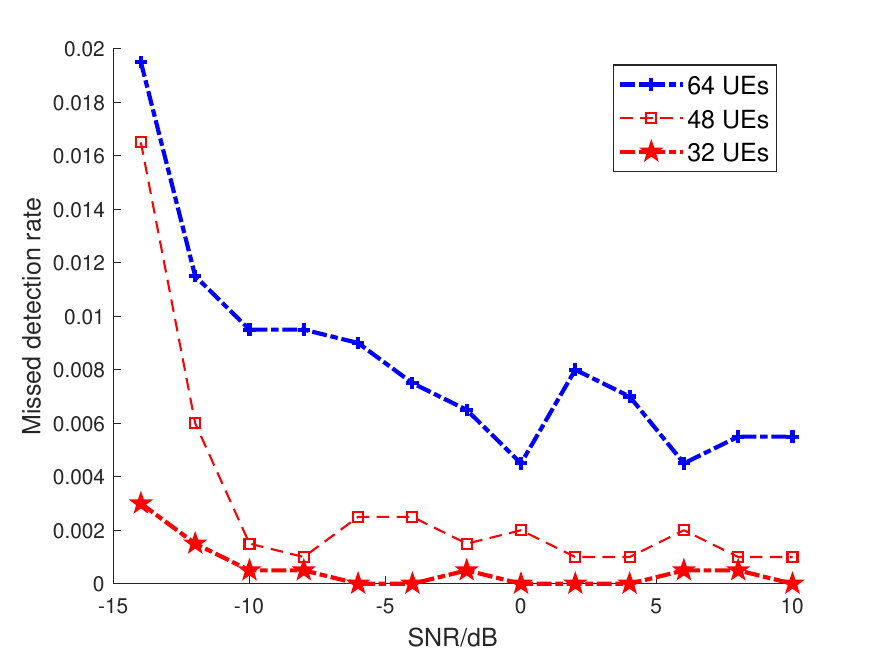}
	\caption{The performance of proposed preamble design.}\label{fig11}
\end{figure}

\begin{figure}[!t]
	\center
	\includegraphics[width=2.4in]{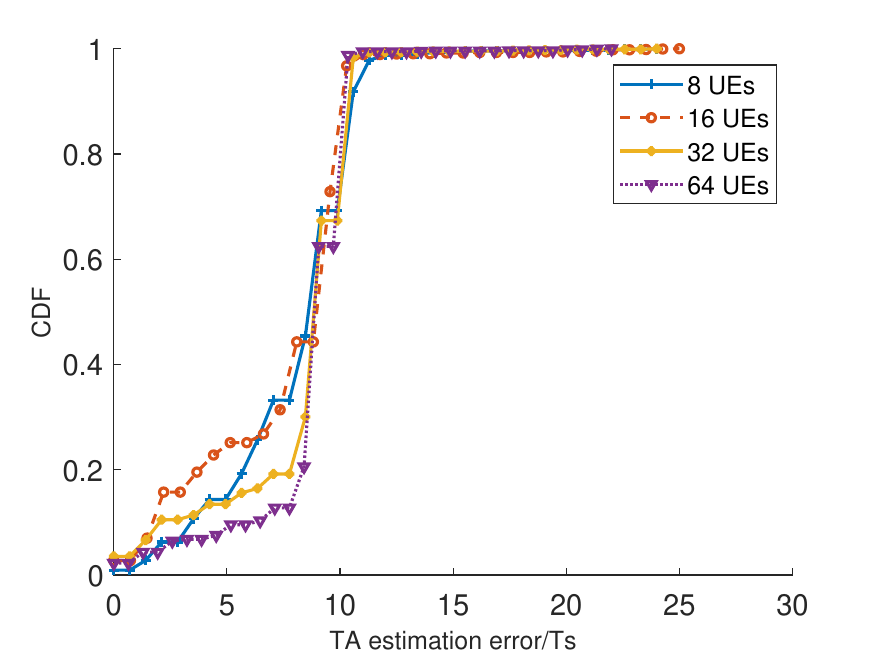}
	\caption{The CDF of TA estimation error.}\label{fig12}
\end{figure}

In Fig.~\ref{fig11}, low missed detection rate performance of our proposal is verified.
It can be seen that the proposed preamble design method supports the access of 64 UEs when SNR is larger than -10 dB.
Here, SNR is the ratio of the strength of the received ZC sequence with a lower amplitude to noise.
Meanwhile, as the number of UEs simultaneously accessing 5G LEO networks increases from 32 to 64, the missed detection rate of preamble decreases slightly.
All curves of missed detection rate decrease rapidly when SNR is less than -4 dB, then tend to be flat and slightly jittering when SNR is large than -4.
That is because the noise term in the received signal is a major obstacle when SNR is lower.
When SNR is larger than -4 dB, inter-preamble interference are dominant and the SINR of UEs almost does not change with the increase of SNR value since the received power of all preambles at the satellite is equal.

In Fig.~\ref{fig12}, we investigate the relationship between TA estimation performance of our proposal and the number of UEs.
Here, we set SNR = -6 dB and other parameters are consistent with the above.
It can be seen that  residual TA estimation error does not exceed the time length of 25 time domain sampling points, regardless of the number of UEs.
Moreover, residual TA estimation errors are mainly limited to 5-10 sampling points.
Note that TA estimation equation (\ref{eq:fractional_TA}) has a rounding operation caused by length difference between time-domain sequence $\{z(l)\}$ and frequency-domain sequence $\{y(n)\}$, and length ratio is $N_{idft}/N_{zc} \approx 7.173$.
Since the TA result in time-domain  is calculated by peak search procedure in frequency domain according to equation (\ref{eq:fractional_TA2}), hence, the TA estimation with around 5-10 sample error means only 1-2 points offset of the optimal $m^*$ estimation in equation (\ref{eq:fractional_TA2}), which represents a high TA estimation accuracy.
Meanwhile, the value of residual TA estimation error is also less that the duration of CP with 288 time domain sampling points when SCS is 30 kHz.
This means our proposed TA estimation approach based on time-frequency pre-compensation and flexible preamble design can well fit into 5G LEO networks.

\begin{figure}[!t]
\center
\includegraphics[width=2.4in]{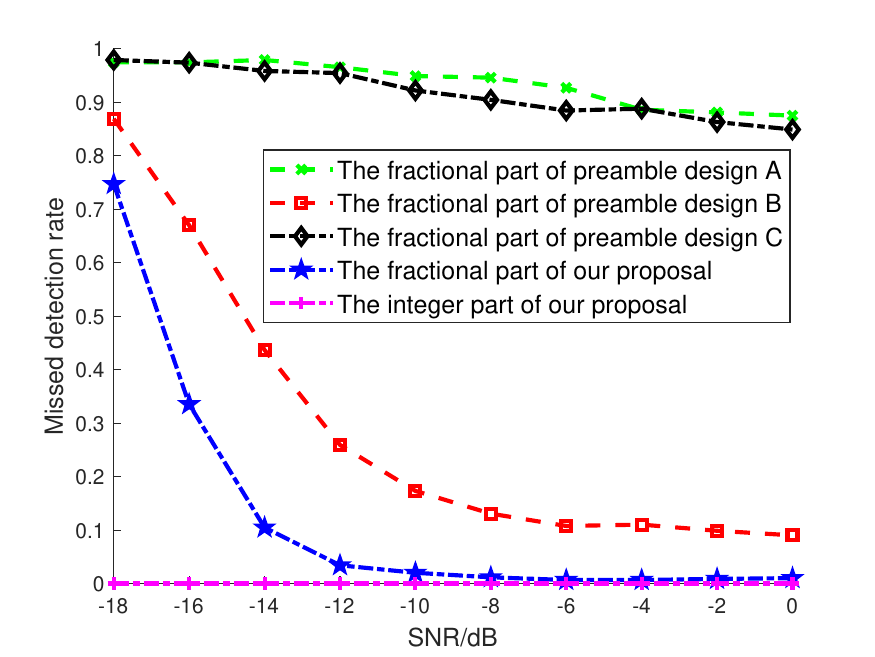}
\caption{The estimation performance of $K_i^u$ and $K_f^u$ under different preamble designs.}\label{fig13}
\end{figure}

\begin{figure}[!t]
\center
\includegraphics[width=2.4in]{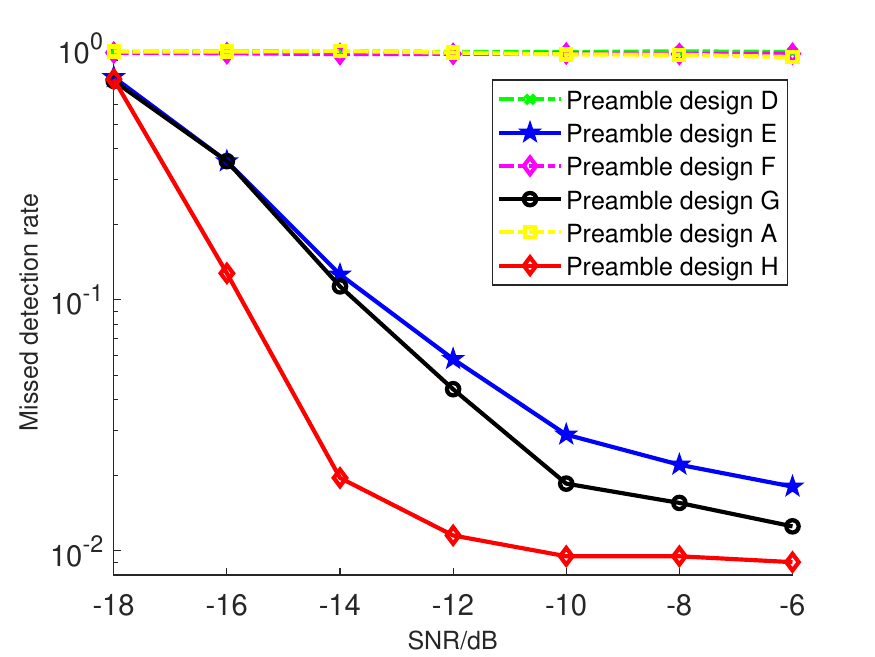}
\caption{The missed detection rate for different preamble designs advised by 3GPP, when 64 UEs access.}\label{fig14}
\end{figure}

Fig.~\ref{fig13} depicts the estimation performance of $K_i^u$ and $K_f^u$ with our proposal and other preamble designs in the scenario with 64-UE access.
It can be seen that, with the increase of SNR, the performance of our proposal, in terms of $K_f^u$ estimation, approaches to that of $K_i^u$ estimation.
Meanwhile, compared with preamble designs A, B and C, our proposal reduces $K_f^u$ estimation error probability by 86.8$\%$, 9.2$\%$, and 84.2$\%$ at SNR = -6 dB, which is because our proposal can mitigate the negative effects of partial-period cross-correlation operations.

In Fig.~\ref{fig14}, when SNR = -6 dB in the scenario with 64-UE access, preamble format designs E, G and H achieve missed detection rate of 1.8$\%$, 1.25$\%$ and 0.9$\%$, respectively.
Meanwhile, preamble format designs A, D and F only achieve missed detection rate of 92.7$\%$, 97.1$\%$ and 95.6$\%$, respectively.
Note that the missed detection rate is defined as the rate that the satellite does not correctly detect a transmitted preamble or estimation error of TA exceeds the duration length of CP of PUSCH.
When 64 UEs access simultaneously, the preamble detection procedure suffers serious inter-preamble interference since it performs large partial-period cross-correlation operations.
In this case, the satellite can detect the presence of preambles, but it cannot obtain exact TA estimation results, which means that UEs access failures.
Finally, we observed that the missed detection rates of existing preamble designs are close to 1, but it does not mean that the transmitted preambles cannot always be detected.
Compared with preamble designs A, D and F, preamble designs E, G and H have significantly improved preamble detection performance, which means the core idea of our preamble format design has high generalization capability and supports multi-UE access in LEO satellite networks.

\section{Conclusions}\label{sec:con}

This article has proposed an enhanced timing advance (TA) estimation approach for random access in low earth orbit satellite networks to improve TA estimation accuracy.
Specifically, to tackle the challenges brought by large frequency offset and arrival time difference among preambles, a time-frequency pre-compensation method at user side has been involved, which helps reduce the missed detection rate of preambles.
Based on the derived upper bound of inter-preamble interference among preamble, it has been concluded that partial-period cross-correlation operations during preamble detection greatly limit the performance of TA estimation.
Under the guidance of the analysis, an improved preamble format design method based on one of format options discussed by 3GPP has been proposed, which features flexible cascading order and differential power allocation of ZC sequences with different roots.
The superiority of our proposal has been verified through both theoretical discussion and numerical results.
By comparing with various baseline schemes, it has been shown that a lower missed detection rate of preamble is obtained and hence TA estimation accuracy can be significantly improved.
In addition, access delay is also a crucial performance metric of random access procedure.
However, since the retransmission strategy design is outside our current research scope, the accurate evaluation of user access delay will be conducted in our future work.


\begin{thebibliography}{99}


\bibitem{integration1}
B. Di, H. Zhang, L. Song, \emph{et al.},
\textquotedblleft Ultra-dense LEO: Integrating terrestrial-satellite networks into 5G and beyond for data offloading,\textquotedblright~
\emph{IEEE Trans. Wireless Commun.}, vol. 18, no. 1, pp. 47-62, Jan. 2019.

\bibitem{integration2}
Y. Sun and M. Peng,
\textquotedblleft Low earth orbit satellite communication supporting direct connection with mobile phones: Key technologies, recent
progress and future directions,\textquotedblright~
\emph{ Telecommunications Science}, vol. 39, no. 2, pp. 25-36, Jan. 2023.

\bibitem{integration3}
Y. Sun, M. Peng, S. Zhang, \emph{et al.},
\textquotedblleft Integrated satellite terrestrial networks: Architectures, key techniques, and experimental
progress,\textquotedblright~
\emph{ IEEE Netw.}, vol. 36, no. 6, pp. 191-198, Dec. 2022.

\bibitem{integration4}
A. Guidotti \emph{et al.},
\textquotedblleft Architectures and key technical challenges for 5G systems incorporating satellites,\textquotedblright~
\emph{IEEE Trans. Veh. Technol.}, vol. 68, no. 3, pp. 2624-2639, Mar. 2019.

\bibitem{integration5}
X. Fang, W. Feng, T. Wei, \emph{et al.}, ``5G embraces satellites for 6G ubiquitous IoT: Basic models for integrated satellite terrestrial networks,'' {\em
IEEE Internet Things J.}, vol. 8, no. 18, pp. 14399-14417, Sep. 2021.

\bibitem{integration6}
O. Kodheli \emph{et al.},
\textquotedblleft Satellite communications in the new space era: A survey and future challenges,\textquotedblright~
\emph{IEEE Commun. Surveys Tuts.}, vol. 23, no. 1, pp. 70-109, Firstquarter 2021.


\bibitem{38.811}
3GPP, \textquotedblleft Study on new radio (NR) to support non terrestrial networks,\textquotedblright~ \emph{3GPP, Technical Report 38.811, v15.4.0}, 2020.

\bibitem{38.821}
3GPP, \textquotedblleft Solutions for NR to support non-terrestrial networks (NTN),\textquotedblright~ \emph{3GPP, Technical Report 38.821, v16.1.0}, 2021.

\bibitem{RA1}
J. -B. Seo and V. C. M. Leung,
\textquotedblleft Performance characterization on random access in LTE-based two-tier small-cell networks,\textquotedblright~
\emph{IEEE Trans. Veh. Technol.}, vol. 65, no. 10, pp. 8528-8537, Oct. 2016.



\bibitem{cell_size}
G. Schreiber and M. Tavares,
\textquotedblleft 5G new radio physical random access preamble design,\textquotedblright~
in \emph{Proc. 2018 IEEE 5G World Forum (5GWF).}, Silicon Valley, CA, USA, Jul. 2018, pp. 215-220.



\bibitem{CFO_damage}
M. Hua, M. Wang, K. W. Yang, \emph{et al.},
\textquotedblleft Analysis of the frequency offset effect on Zadoff--Chu sequence timing performance,\textquotedblright~
\emph{IEEE Trans. Commun.}, vol. 62, no. 11, pp. 4024-4039, Nov. 2014.

\bibitem{beam_size}
A. Guidotti,
\textquotedblleft Beam size design for new radio satellite communications systems,\textquotedblright~
\emph{IEEE Trans. Veh. Technol.}, vol. 68, no. 11, pp. 11379-11383, Nov. 2019.

\bibitem{38.221}
3GPP, \textquotedblleft NR; Physical channels and modulation,\textquotedblright~ \emph{3GPP, Technical specification 38.221, v17.1.0}, 2022.

\bibitem{GNSS_practice}
O. Kodheli \emph{et al.},
\textquotedblleft Random access procedure over non-terrestrial networks: From theory to practice,\textquotedblright~
\emph{IEEE Access.}, vol. 9, pp. 109130-109143, 2021.



\bibitem{R19}
3GPP, \textquotedblleft New SID: Study on satellite access - Phase 3,\textquotedblright~
\emph{3GPP TSG RAN WG1 $\#$99 SP-220679}, Jun. 7, 2022. [Online]. Available: {\url{https://www.3gpp.org/ftp/Information/WI_Sheet/SP-220679.zip}}.


\bibitem{location}
W. Wang, T. Chen, R. Ding, \emph{et al.},
\textquotedblleft Location-based timing advance estimation for 5G integrated LEO satellite communications,\textquotedblright~
\emph{IEEE Trans. Veh. Technol.}, vol. 70, no. 6, pp. 6002-6017, Jun. 2021.


\bibitem{beam_subarea}
C. Li, H. Ba, H. Duan, \emph{et al.}, \textquotedblleft A two-step time delay difference estimation method for initial random access in satellite LTE system,\textquotedblright~
\emph{in Proc. 16th Int. Conf. Adv. Commun. Technol.}, Pyeongchang, Korea (South), 2014, pp. 10-13.

\bibitem{zhenli}
L. Zhen, H. Qin, B. Song, \emph{et al.},
\textquotedblleft Random access preamble design and detection for mobile satellite communication systems,\textquotedblright~
\emph{IEEE J. Sel. Areas Commun.}, vol. 36, no. 2, pp. 280-291, Feb. 2018.

\bibitem{COE}
H. Chougrani, S. Kisseleff, W. A. Martins, \emph{et al}., \textquotedblleft NB-IoT random access for Nonterrestrial networks: Preamble detection and uplink synchronization,\textquotedblright~ {\em
IEEE Internet Things J.}, vol. 9, no. 16, pp. 14913-14927, Aug. 2022.


\bibitem{weighted}
C. E. M. Silva, F. J. Harris and G. J. Dolecek,
\textquotedblleft Synchronization algorithms based on weighted CAZAC preambles for OFDM systems,\textquotedblright~
in \emph{Proc. 2013 13th International Symposium on Communications and Information Technologies (ISCIT)}, Surat Thani, Thailand, Sep. 2013, pp. 192-197.




\bibitem{2NR-preamble}
T. A. Khan and X. Lin,
\textquotedblleft Random access preamble design for 3GPP non-terrestrial networks,\textquotedblright~
in \emph{Proc. 2021 IEEE Globecom Workshops (GC Wkshps)}, Madrid, Spain, Dec. 2021, pp. 1-5.

\bibitem{2root}
C. Zhang, W. Cao, N. Zhang, \emph{et al.},
\textquotedblleft Root pair selection for two-root random access preamble,\textquotedblright~
in \emph{Proc. 2021 IEEE 93rd Vehicular Technology Conference (VTC2021-Spring)}, Helsinki, Finland, Apr. 2021, pp. 1-6.


\bibitem{ZC_decomposed}
Y. J. Kim, M. Asim and Y. S. Cho,
\textquotedblleft Preamble design technique for accurate timing/positioning in high doppler environments,\textquotedblright~
\emph{IEEE Trans. Veh. Technol.}, vol. 71, no. 6, pp. 6784-6789, Jun. 2022.




\bibitem{one-step1}
L. Zhen, T. Sun, G. Lu, \emph{et al.},
\textquotedblleft Preamble design and detection for 5G enabled satellite random access,\textquotedblright~
\emph{IEEE Access.}, vol. 8, pp. 49873-49884, 2020.



\bibitem{one-step2}
L. Zhen, A. K. Bashir, K. Yu, \emph{et al.},
\textquotedblleft Energy-efficient random access for LEO satellite-assisted 6G internet of remote things,\textquotedblright~
\emph{IEEE Internet Things J.}, vol. 8, no. 7, pp. 5114-5128, Apr. 2021.


\bibitem{Multisatellite}
B. Zhao, G. Ren and H. Zhang,
\textquotedblleft Multisatellite cooperative random access scheme in low earth orbit satellite networks,\textquotedblright~
\emph{IEEE Internet Things J.}, vol. 13, no. 3, pp. 2617-2628, Sep. 2019.


\bibitem{hua1}
M. Hua \emph{et al.},
\textquotedblleft Analysis of the frequency offset effect on random access signals,\textquotedblright~
\emph{IEEE Trans. Commun.}, vol. 61, no. 11, pp. 4728-4740, Nov. 2013.


\bibitem{chen_shanzhi}
S. Chen, S. Sun and S. Kang,
\textquotedblleft System integration of terrestrial mobile communication and satellite communication - the trends, challenges and key technologies in B5G and 6G,\textquotedblright~
\emph{Commun., China.}, vol. 17, no. 12, pp. 156-171, Dec. 2020.

\bibitem{damange_O}
X. Lin, S. Cioni, G. Charbit, \emph{et al.},
\textquotedblleft On the path to 6G: Embracing the next wave of low earth orbit satellite access, \textquotedblright~
\emph{IEEE Commun. Mag.}, vol. 59, no. 12, pp. 36-42, Dec. 2021.

\bibitem{damange_LO}
Y. Cao and T. Zhang,
\textquotedblleft Two stage frequency offset pre-compensation scheme for satellite mobile terminals,\textquotedblright~
in \emph{Proc. 2018 13th IEEE Conference on Industrial Electronics and Applications (ICIEA).}, Wuhan, China, Jun. 2018, pp. 117-122.



\bibitem{huawei}
3GPP, \textquotedblleft Discussion on doppler compensation, timing advance and RACH for NTN,\textquotedblright~ Huawei, HiSilicon, Reno, USA,
\emph{3GPP TSG RAN WG1 $\#$99 R1-1911860}, Nov. 18, 2019. [Online]. Available: {\url{https://www.3gpp.org/ftp/tsg_ran/WG1_RL1/TSGR1_99/Docs/R1-1911860.zip}}.


\bibitem{f_est1}
J. Khalife, M. Neinavaie and Z. M. Kassas,
\textquotedblleft Blind doppler tracking from OFDM signals transmitted by broadband LEO satellites,\textquotedblright~
in \emph{Proc. Veh. Technol. Conf}, Helsinki, Finland, Apr. 2021, pp. 1-5.


\bibitem{f_est2}
W. Wang, Y. Tong, L. Li, \emph{et al.},
\textquotedblleft Near optimal timing and frequency offset estimation for 5G integrated LEO satellite communication system,\textquotedblright~
\emph{IEEE Access.}, vol. 7, pp. 113298-113310, 2019.

\bibitem{f_est3}
D. Tian, Y. Zhao, J. Tong, \emph{et al.},
\textquotedblleft Frequency offset estimation for 5G based LEO satellite communication systems,\textquotedblright~
in \emph{Proc. IEEE/CIC International Conference on Communications in China (ICCC)}, Changchun, China, Aug. 2019, pp. 647-652.


\bibitem{Least-Squares}
J. Yan, C. C. J. M. Tiberius, P. J. G. Teunissen, \emph{et al.},
\textquotedblleft A framework for low complexity least-squares localization with high accuracy,\textquotedblright~
\emph{IEEE Trans. Signal Process.}, vol. 58, no. 9, pp. 4836-4847, Sep. 2010.


\bibitem{position_error}
R. Morales-Ferre, E. S. Lohan, G. Falco, \emph{et al.},
\textquotedblleft GDOP-based analysis of suitability of LEO constellations for future satellite-based positioning,\textquotedblright~
in \emph{Proc. IEEE Int. Conf. Wireless Space Extreme Environ (WiSEE)}, Vicenza, Italy, Oct. 2020, pp. 147-152.


\bibitem{matrix}
X. Chen, M. Wang and L. Zhang,
\textquotedblleft Analysis on the performance bound of doppler positioning using one LEO satellite,\textquotedblright~
in \emph{Proc. Veh. Technol. Conf.}, Nanjing, China, May. 2016, pp. 1-5.


\bibitem{Partial-Period}
T. Lee and K. Yang,
\textquotedblleft Partial-period correlations of Zadoff-Chu sequences and their relatives,\textquotedblright~
\emph{IEEE Trans. Inf. Theory.}, vol. 60, no. 9, pp. 5791-5802, Sep. 2014.


\bibitem{windowing_sequence}
K. G. Paterson and P. J. G. Lothian,
\textquotedblleft Bounds on partial correlations of sequences,\textquotedblright~
\emph{IEEE Trans. Inf. Theory.}, vol. 44, no. 3, pp. 1164-1175, May. 1998.


\bibitem{Theory}
J. Stillwell, \emph{Elements of Number Theory}. New York, NY, USA: Springer-Verlag, 2002.

\bibitem{zte}
3GPP, \textquotedblleft Discussion on the TA and PRACH for NTN,\textquotedblright~ ZTE, Reno, USA,
\emph{3GPP TSG RAN WG1 $\#$99 R1-1912612}, Nov. 18, 2019. [Online]. Available: {\url{https://www.3gpp.org/ftp/tsg_ran/WG1_RL1/TSGR1_99/Docs/R1-1912612.zip}}.

\bibitem{metric}
M. Caus, A. I. P¨¦rez-Neira, J. Bas, \emph{et al.},
\textquotedblleft New satellite random access preamble design based on pruned DFT-spread FBMC,\textquotedblright~
\emph{IEEE Trans. Commun.}, vol. 68, no. 7, pp. 4592-4604, Jul. 2020.

\bibitem{common_TA3}
H. Saarnisaari, A. O. Laiyemo and C. H. M. de Lima,
\textquotedblleft Random access process analysis of 5G new radio based satellite links,\textquotedblright~
in \emph{Proc. IEEE 2nd 5G World Forum}, Dresden, Germany, Sep. 2019, pp. 654-658.

\bibitem{missed_detection_rate2}
H. Chougrani, S. Kisseleff and S. Chatzinotas,
\textquotedblleft Efficient preamble detection and time-of-arrival estimation for single-tone frequency hopping random access in NB-IoT,\textquotedblright~
{\em
IEEE Internet Things J.}, vol. 8, no. 9, pp. 7437-7449, May, 2021.



\end{thebibliography}
\end{document}